\documentclass[journal]{IEEEtran}

\usepackage{cite}
\usepackage{array}
\usepackage{color}
\usepackage{float}
\usepackage[cmex10]{amsmath}
\usepackage{nomencl}						
\usepackage[normalem]{ulem}			        
\usepackage{diagbox}						
\usepackage{slashbox}						
\usepackage{colortbl}						
\usepackage{multirow}						
\usepackage{tabularx}
\usepackage{placeins}
\usepackage{algorithm}
\usepackage{mathrsfs}
\usepackage{mathdots}
\usepackage{amssymb}
\usepackage{dblfloatfix}
\usepackage{amsthm}
\usepackage{amsmath}
\usepackage{arydshln}
\usepackage{color}
\usepackage[noend]{algpseudocode}
\usepackage{bm}
\usepackage{url}
\usepackage{threeparttable}
\usepackage[hidelinks]{hyperref}
\usepackage[T1]{fontenc}
\newcommand{\subparagraph}{}
\usepackage{fancyhdr} 
\usepackage{extarrows}
\usepackage{enumitem}
\usepackage{booktabs}
\usepackage{threeparttable}

\usepackage{booktabs,siunitx,makecell}
\usepackage{pifont}

\ifCLASSINFOpdf
  \usepackage[pdftex]{graphicx}
	\graphicspath{{./figure/}}
  \DeclareGraphicsExtensions{.pdf,.jpeg,.png}
\else
  \usepackage[dvips]{graphicx}
  \graphicspath{{./figure/}}
  \DeclareGraphicsExtensions{.eps}
\fi

\ifCLASSOPTIONcompsoc
  \usepackage[caption=false,font=normalsize,labelfont=sf,textfont=sf]{subfig}
\else
  \usepackage[caption=false,font=footnotesize]{subfig}
\fi

\newcommand{\figref}[1]{\figurename~\ref{#1}}
\newcommand{\tabref}[1]{Table~\ref{#1}}

\ifCLASSOPTIONonecolumn

\else

\renewcommand{\arraystretch}{1.5}
\fi

\begin{document}
\setcounter{page}{1}
\vspace{-2\baselineskip}
\title{Consideration of Control-Loop Interaction in Transient Stability of Grid-Following Inverters using Bandwidth Separation Method}
\author{Yifan~Zhang, \IEEEmembership{Student Member, IEEE}, Yunjie~Gu, \IEEEmembership{Senior Member, IEEE}, Yue~Zhu, \IEEEmembership{Member, IEEE}, Yitong~Li, \IEEEmembership{Member, IEEE}, Hsiao-Dong Chiang, \IEEEmembership{Fellow, IEEE}, Timothy~C.~Green, \IEEEmembership{Fellow, IEEE} 
\thanks{This work was supported by the Royal Society under the award ICA{\textbackslash}R2{\textbackslash}242103 and the Ralph O'Connor Sustainable Energy Institute at Johns Hopkins University.
Yifan Zhang, Yunjie Gu, Yue Zhu, and Timothy~C.~Green are with Imperial College, London, UK, Yitong Li is with the State Key Laboratory of Electrical Insulation and Power Equipment, School of Electrical Engineering, Xi'an Jiaotong University, Xi'an 710049, China, and Hsiao-Dong~Chiang is with 
Cornell University, Ithaca, NY, US.
(e-mails: 
\url{yifan.zhang21@imperial.ac.uk};
\url{yunjie.gu@imperial.ac.uk};
\url{yue.zhu18@imperial.ac.uk};
\url{t.green@imperial.ac.uk};
\url{yitongli@xjtu.edu.cn};
\url{hc63@cornell.edu}.
Corresponding author: Yunjie Gu.)
}
}
\IEEEaftertitletext{\vspace{-2.5\baselineskip}}
\ifCLASSOPTIONpeerreview
	\maketitle 
\else
	\maketitle
\fi
\thispagestyle{fancy}
\lhead{IEEE TRANSACTIONS ON POWER ELECTRONICS}
\rhead{\thepage}
\cfoot{}
\renewcommand{\headrulewidth}{0pt}
\pagestyle{fancy}

\begin{abstract}
Grid-following inverters have been widely adopted as a grid interface of renewable energy and ensuring their small- and large-signal stability is critical to modern power systems. Their large-signal, or transient, stability is a significant challenge to analyze because of the interaction of the phased-locked loop (PLL) that must maintain synchronism with various outer loop controllers with whatever outer-loop controllers exist. Simple analysis in which outer loop controller are idealized is not sufficient and interactions of the non-linear dynamics of PLL with the dynamics of the DC-link voltage control (DVC) and, where present, the AC terminal voltage control (TVC). An asymptotic analysis approach, termed \textit{bandwidth separation method}, is proposed. This method enables simplification and order reduction of the original differential equations when sufficient bandwidth separation exists. Through this method, the interaction between the DVC and PLL is explicitly characterized, revealing that such interaction degrades system stability and shrinks the stability region. The analysis also indicates that voltage instability, rather than PLL loss of synchronization alone, is often the root cause of transient instability. Optimal bandwidth configurations for PLL and DVC are identified under various grid fault conditions: a larger PLL bandwidth improves resilience to phase jump faults, while a larger DVC bandwidth enhances tolerance to power fluctuations. In addition, the influence of the TVC loop is analyzed, showing that a high TVC bandwidth can mitigate the destabilizing effects of PLL–DVC interaction and further improve transient stability. All analytical findings are validated through hardware-in-the-loop (HIL) experiments.
\end{abstract}


\section{Introduction} \label{section_intro}
The high penetration of renewable energy and large-scale application of inverter-based resources (IBRs) — such as wind farms, photovoltaic plants, and HVDC circuits — introduce complex and diverse dynamic behaviors under various disturbances, posing significant challenges to the stability and control of modern power systems \cite{1705631}. Transient stability analysis of power systems considering IBRs has become an important topic in electrical power engineering \cite{Gu_proceeding, Xiongfei_overview}. Grid-following (GFL) inverters, which use a phase-locked loop (PLL) for synchronization with the grid, have been widely adopted as a grid interface of renewable energy, and ensuring the transient stability of GFL inverters has grown as a challenge, especially in weak grids \cite{Gu_proceeding, Xiongfei_overview, 8632731, hu2019large}. 

Modeling of PLLs and analysis of transient stability have been well developed in \cite{8632731, fu2020large, hu2019large, he2019transient, wu2019design, ma2021generalized, zhao2020nonlinear}. The dynamics of a PLL can be described as a second-order nonlinear differential equation. However, much of the research undertaken on GFL inverters focuses exclusively on the dynamic process of the PLL and overlooks other control loops, such as DC-link voltage control (DVC) and AC-side terminal voltage control (TVC), and the interactions that can occur between these control loops. 

For small-signal stability, the analysis of these three outer control loops (PLL, DVC, TVC, where current control is an inner loop) has been well-studied, and references \cite{zhou2014stability, huang2017effect, wu2020impact}, for example, analyze the interaction among loops and the influence of loop bandwidths on stability. In contrast, large-signal analysis involving multiple control loops has only been covered in a few papers \cite{ZiqianY, TaulM, liu2022impact, PalD, 10785310, wang2023transient, priyamvada2020online}. Reference \cite{PalD, 10785310} modifies the differential equation of a PLL to incorporate the impact of the DVC loop on PLL, but the dynamics of the DVC loop itself are not considered. Although the instability caused by input power variation is discussed in \cite{10567400}, the influence of control parameters and the underlying instability mechanism are not analyzed. Reference \cite{liu2022impact} investigates the effect of DVC parameters using phase portraits from simulation studies, but focuses solely on relative stability without examining the impact on the stability region. In \cite{wang2023transient, priyamvada2020online}, the instability of the DVC control loop is studied; however, the PLL dynamics and their interactions with the DVC loop are neglected. Nevertheless, these studies collectively indicate that transient instability cannot be solely attributed to the loss of synchronization of the PLL. In \cite{TaulM, ZiqianY}, GFL inverters with all three outer control loops are modeled, leading to fifth-order nonlinear ordinary differential equations (ODEs) that describe the system dynamics. Due to the difficulty of performing theoretical analysis on high-order nonlinear ODEs, all these works rely on numerical methods for stability analysis. An analytical simplification approach for high-order inverter models has yet to be established.

To address this issue, a \textit{bandwidth separation method} was first proposed for GFL inverters in our previous work \cite{previouswork} to reveal the instability mechanism of the GFL inverter considering multiple control loops. In this paper, the proposed \textit{bandwidth separation method} is further extended to a more general range of parameter configurations, and a more detailed analysis is presented. Under various categories of bandwidth settings, the proposed method yields a reduced model that makes it easier to compute the region of attraction (ROA) and visualize it in the phase plane. A broader set of cases, including various large-signal disturbances, is considered, and hardware-in-the-loop (HIL) experiments are conducted to validate the effectiveness of the proposed analysis approach.

Unlike the conventional method directly focusing on the coupling effects of different control loops, the proposed method first intentionally separates the bandwidths of the outer control loops. This separation enables a clearer understanding of how bandwidth configurations influence transient stability and provides insight into the interaction mechanisms among the loops. Using this method, the stability region can be readily derived. Furthermore, this paper illustrates in detail how the interaction between the DVC and PLL degrades system stability, revealing a bandwidth-dependent trade-off between phase-angle stability and dc-link stability. This paper also shows that a sufficiently high TVC bandwidth improves stability: it expands the stability region for both the phase-angle and dc-link dynamics and attenuates the adverse interaction between the PLL and DVC. In addition, the optimal combination of PLL and DVC bandwidths is thereby identified for various types of disturbance.

\section{The Proposed Bandwidth Separation Method}
\subsection{Modelling of a GFL inverter}
\begin{figure}[t!]
\centering
\subfloat[Overall control system structure.]{\includegraphics[width=0.48\textwidth]{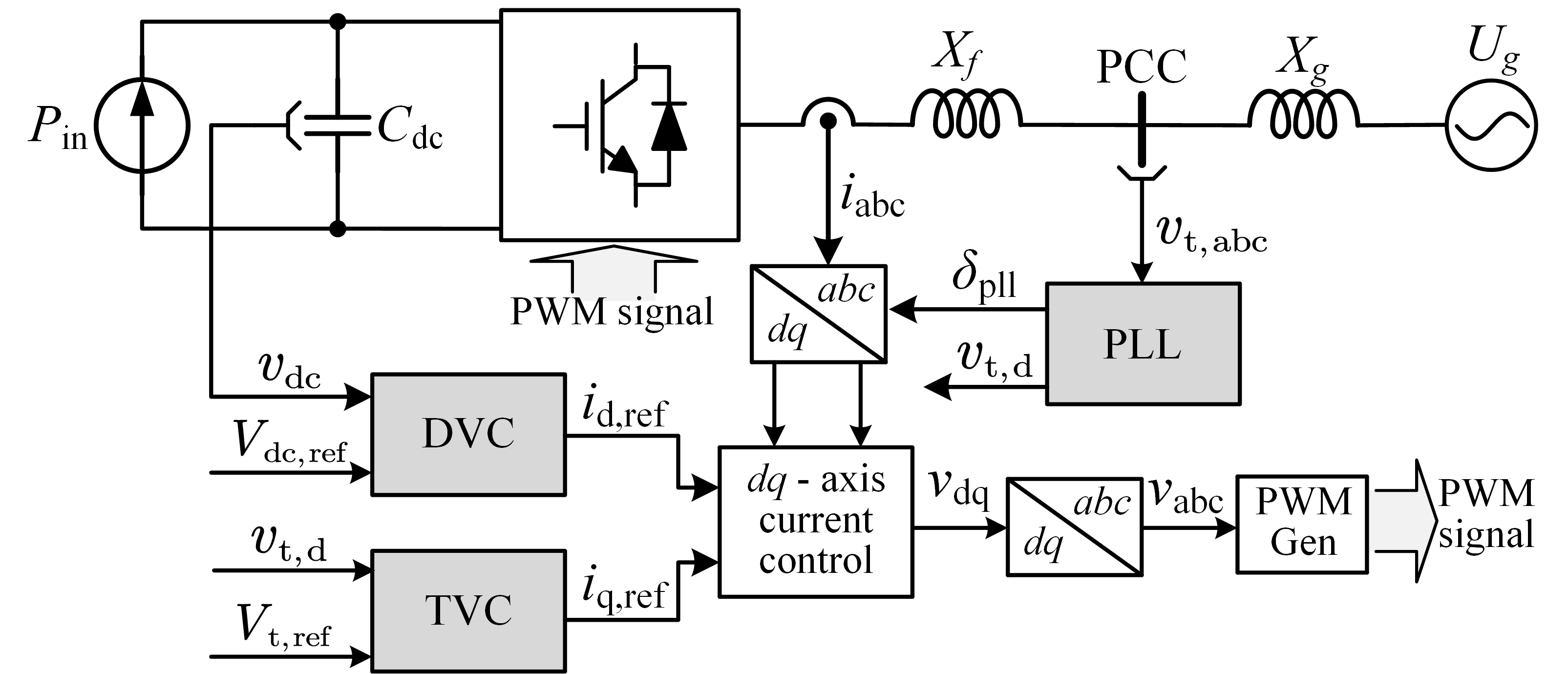}} \\
\vspace{-0.5mm}
\subfloat[Phase-locked loop (PLL).]{\includegraphics[width=0.33\textwidth]{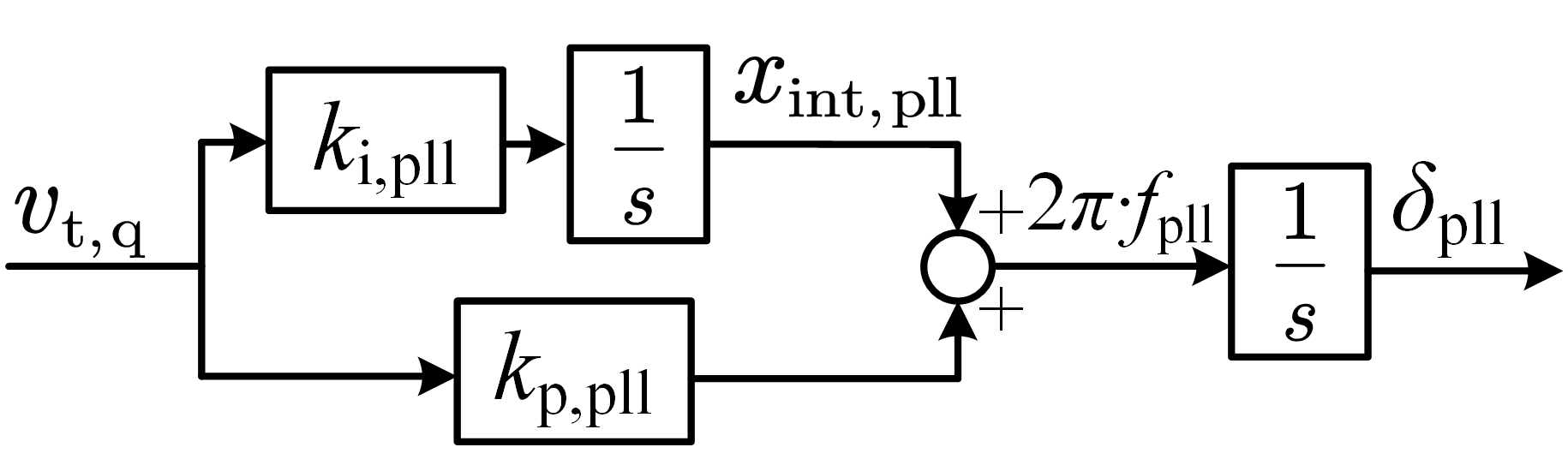}}\\
\vspace{-0.5mm}
\subfloat[DC-link voltage control (DVC).]{\includegraphics[width=0.34\textwidth]{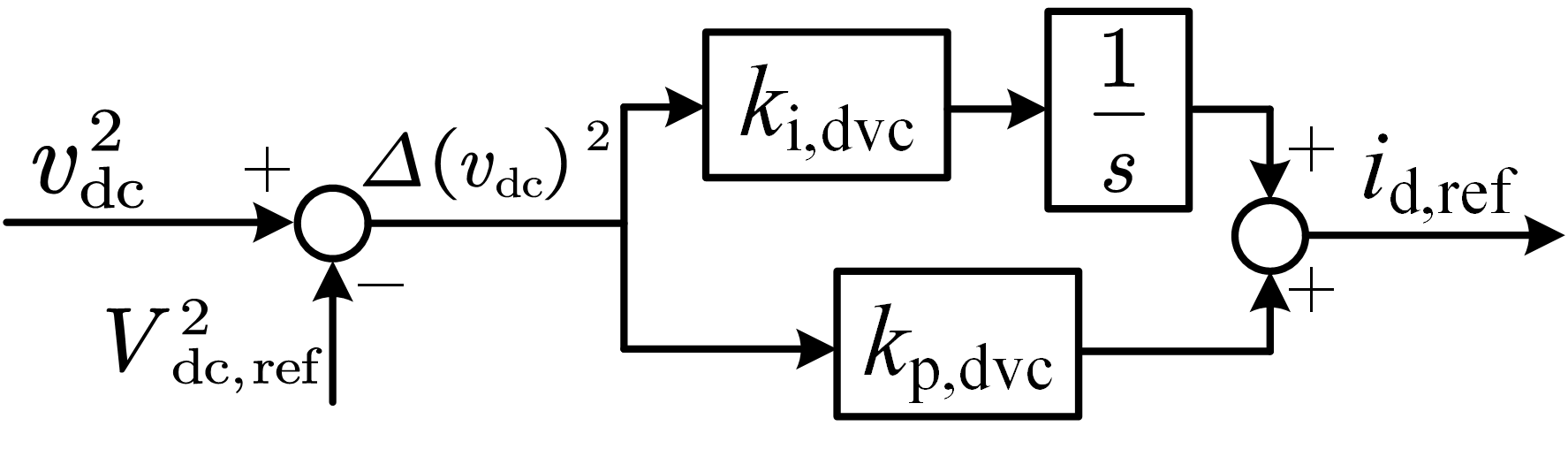}}\\
\vspace{-0.5mm}
\subfloat[Terminal voltage control (TVC).]{\includegraphics[width=0.29\textwidth]{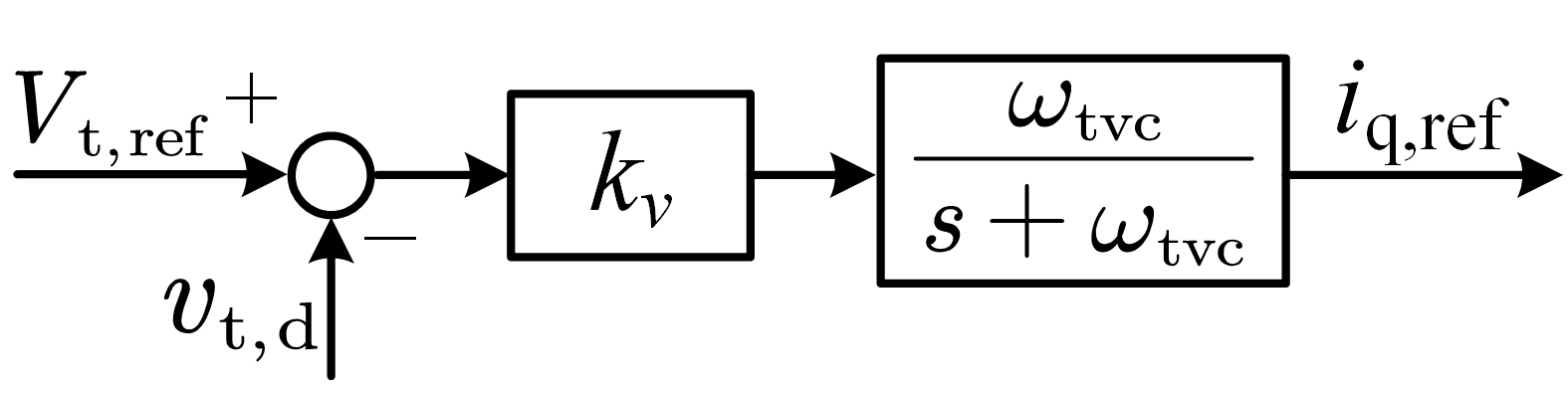}}
\caption{System diagram of grid-following inverter.}
\label{GFL}
\end{figure}
The diagram of a single-machine infinite-bus system and control loops in the GFL inverter, including PLL, DVC, and TVC is shown in \figref{GFL}~(a). The grid is represented by $U_g$ with fundamental frequency $\omega_s$. In general, the control bandwidth of any of the outer loops is much less than that of the inner current loop, and thus a grid-following inverter can be modeled as a controlled current source (that is, $i_d=i_{\text{d,ref}}$, $i_q=i_{\text{q,ref}}$) when analyzing its transient dynamics \cite{TaulM, Xiongfei_overview}. Typical formats of outer loop control will now be discussed. As for the PLL, a proportional-integral (PI) controller is used to drive the observed $q$-axis voltage $v_{\text{t,q}}$ to zero, as shown in \figref{GFL}~(b). The nonlinear dynamics can be expressed as follows:
\vspace{-0.5mm}
\begin{equation}
\begin{cases}
	\frac{d}{dt}\delta =k_{\mathrm{p},\mathrm{pll}}v_{\mathrm{t},\mathrm{q}}+x_{\mathrm{int},\mathrm{pll}}\\
	\frac{d}{dt}x_{\mathrm{int},\mathrm{pll}}=k_{\mathrm{i},\mathrm{pll}}v_{\mathrm{t},\mathrm{q}}\\
	v_{\mathrm{t},\mathrm{q}}=X_gi_d+i_qR_g-U_g\sin \delta +i_d\dot{\delta}L_g/\omega _s\\
\end{cases}
\label{PLL}
\end{equation}
where ${x}_{\mathrm{int},\mathrm{pll}}$ is the integral term in the PI controller and $\delta$ represents the angle difference between the PLL and the grid.

The DC-link voltage control (DVC) is illustrated in \figref{GFL}~(c). The variable $\varDelta \left( v_{dc} \right) ^2$ is defined as the deviation between the square of the actual DC-link voltage and its reference, given by $v_{\mathrm{dc}}^{2} - V_{\mathrm{dc,ref}}^{2}$, and PI controller is then employed to regulate the voltage by adjusting the $d$-axis current reference \cite{PalD, Xiongfei_overview}. The nonlinear dynamics can be expressed as follows:
\vspace{-0.5mm}
\begin{equation}
\begin{cases}
	\frac{d}{dt}i_d=k_{\mathrm{i},\mathrm{dvc}}\varDelta \left( v_{\text{dc}} \right) ^2+k_{\mathrm{p},\mathrm{dvc}}\frac{2}{C_{dc}/\omega _s}\left( P_{in}-p \right)\\
	\frac{d}{dt}\varDelta \left( v_{\text{dc}} \right) ^2=\frac{2}{C_{\mathrm{dc}}/\omega _s}\left( P_{in}-p \right)\\
	p={v_{\mathrm{t},\mathrm{dq}}}^{\intercal}\cdot i_{\mathrm{dq}}=i_dU_g\cos \delta -i_qU_g\sin \delta +i_{dq}^{2}R_g\\
\end{cases}
\label{DVC}
\end{equation}

In terms of TVC, droop control with a low-pass filter is typically employed to regulate the injection of reactive current according to grid codes \cite{zhao2020reactive, netz2006grid, tenne2012requirements}, as shown in \figref{GFL}~(d). This control approximates the standard $v_t\text{-}i_q$ droop characteristic, assuming the PLL angle is aligned with the grid voltage such that $v_{\text{t,d}} \approx v_t$, where $v_t$ is the terminal voltage measured at the PCC in \figref{GFL}~(d). In this figure, $\omega_{\mathrm{tvc}}$  denotes the cut-off frequency of the low-pass filter, and $k_v$ is the droop coefficient. The nonlinear dynamics can be expressed as follows:
\begin{equation}
\left\{ \begin{array}{l}
	\frac{1}{\omega _{\mathrm{tvc}}} \cdot\frac{d}{dt}i_q=-i_q+\left( v_{\mathrm{t},\mathrm{d}}-V_{\mathrm{t},\mathrm{ref}} \right) \cdot k_v\\
	v_{\mathrm{t},\mathrm{d}}=U_g\cos \delta +i_dR_g-i_qX_g+i_q\dot{\delta}L_g/\omega _s\\
\end{array} \right. 
\label{TVC}
\end{equation}
Equations \eqref{PLL}, \eqref{DVC}, and \eqref{TVC} can be rewritten in ordinary differential equation (ODE) format as:

\footnotesize
\begin{subequations}
\begin{equation}
\frac{1}{\omega _{\mathrm{pll}}}\cdot \frac{d}{dt}\left[ \begin{array}{c}
	\delta\\
	x_{\mathrm{int},\mathrm{pll}}\\
\end{array} \right] =\left[ \begin{array}{c}
	f_{\mathrm{pll},1}\left( \delta ,i_d,\frac{x_{\mathrm{int},\mathrm{pll}}}{\omega _{\mathrm{pll}}} \right)\\
	f_{\mathrm{pll},2}\left( \delta ,i_d,x_{\mathrm{int},\mathrm{pll}} \right)\\
\end{array} \right]  
\label{ODE1}
\end{equation}
\vspace{-0.5mm}
\begin{equation}
\frac{1}{\omega _{\mathrm{dvc}}}\cdot \frac{d}{dt}\left[ \begin{array}{c}
	i_d\\
	\omega _{\mathrm{dvc}}\varDelta \left( v_{dc} \right) ^2\\
\end{array} \right] =\left[ \begin{array}{c}
	f_{\mathrm{dvc},1}\left( \delta ,i_{\mathrm{dq}},\omega _{\mathrm{dvc}}\varDelta \left( v_{dc} \right) ^2 \right)\\
	f_{\mathrm{dvc},2}\left( \delta ,i_{\mathrm{dq}} \right)\\
\end{array} \right] 
\label{ODE2}
\end{equation}
\vspace{-0.5mm}
\begin{equation}
\frac{1}{\omega _{\mathrm{tvc}}}\cdot \frac{d}{dt}i_q=f_{\mathrm{tvc}}\left( \delta ,i_q \right)  
\label{ODE3}
\end{equation}
\label{ODE}
\end{subequations}
\normalsize
in which the resistance of the grid impedance has been treated as negligible.

In the system defined by \eqref{ODE}, the control bandwidths of the PLL, DVC, and TVC loops are denoted by $\omega_{\mathrm{pll}}$, $\omega_{\mathrm{dvc}}$, and $\omega_{\mathrm{tvc}}$, respectively. Their values are given by  $\omega_{\mathrm{pll}}=k_{\mathrm{p,pll}}$, $\omega _{\mathrm{dvc}}=\frac{2}{C_{\mathrm{dc}}/\omega _s}\cdot k_{\mathrm{p},\mathrm{dvc}}$. The derivation of the bandwidth expressions, the controller designs and the detailed mathematical expressions for all functions appearing in \eqref{ODE} are provided in Appendix~\ref{A}.

Eq. \eqref{ODE} constitutes the full-order model of the outer loop control system and it can be seen that it is $5^{th}$ order, which is relatively high, and includes several nonlinear terms introduced by the PLL, the power calculation, and the $dq$ transform. Thus, it is not straightforward to analyze the transient stability of the full system model and an appropriate approach is required to simplify and reduce the order of the ODE model.

\subsection{Bandwidth Separation Method}
The proposed method is underpinned by singular perturbation theory \cite{smith1985singular, nonlinearbook_Khalil, o1988nonlinear}, which enables the decomposition of a dynamical system into slow and fast subsystems under sufficient time-scale separation. A standard singularly perturbed system takes the form:
\begin{equation}
\begin{aligned}
       \dot{\mathbf{X}}=f(\mathbf{X},\mathbf{Y},\varepsilon ) \\
        \varepsilon \dot{\mathbf{Y}}=g(\mathbf{X},\mathbf{Y},\varepsilon ) 
\end{aligned}
    \label{ODE_o}
\end{equation}
where $\varepsilon$ is a small positive parameter, and $\mathbf{X}$ and $\mathbf{Y}$ denote slow and fast variables, respectively. As $\varepsilon \to 0$, the system reduces to a differential-algebraic form:
\begin{equation}
\begin{aligned}
    \dot{\mathbf{X}} = f(\mathbf{X},\mathbf{Y},0 ) \\
    \mathbf{0} = g(\mathbf{X},\mathbf{Y},0 )
\end{aligned}
\label{DAE_o}
\end{equation}
According to Tikhonov’s theorem (Theorem 11.2 in~\cite{nonlinearbook_Khalil}), if the fast subsystem is exponentially stable for fixed $\mathbf{X}$, the solution $\mathbf{Y}$ quickly converges to the algebraic constraint $g(\mathbf{X}, \mathbf{Y}, 0) = \mathbf{0}$, and the reduced-order differential-algebraic equation (DAE) approximates the full system with an $O(\varepsilon)$ error. In addition, Theorem 16-8 in~\cite{chiang2015stability} shows that the stability boundary of the full system can be approximated by combining the stability region of the slow subsystem with the union of stable manifolds of the fast subsystem. This implies that if both subsystems are individually stable, the full system is also stable~\cite{kuehn2015chaos}.

It can be seen that the ODE model represented by \eqref{ODE} has the explicit singular perturbation structure of \eqref{ODE_o}. When the bandwidths of the three outer loops of a GFL inverter are very different, the singular perturbation theory can be applied to can be applied to separate the system into fast and slow subsystems, enabling the derivation of a reduced-order DAE model to capture the system dynamics. In this framework, the control loops with the larger bandwidths can be regarded as the fast subsystem, and the reciprocal of this bandwidth serves as the time-scale parameter $\varepsilon$ in \eqref{ODE_o} of the fast subsystem. Loops whose reciprocal bandwidths are $O(\varepsilon)$ or smaller are included in the fast subsystem, while the remaining loops constitute the slow subsystem.

A time-domain simulation comparison is carried out in Section III, which demonstrates that if the bandwidths of outer control loops in GFL inverters differ by a factor of 7-8, then a good approximation of the DAE model can be obtained. Such bandwidth separation may be prevalent in the industry so that the loops can be tuned independently of each other. 

The analysis steps in the next Section are as follows: 
\begin{enumerate}
    \item Determine the stability region, $\Omega_f$ of the fastest control loop. At this stage, the variables in the other two loops are treated as slow variables and approximated as constants. 

    \item Determine the stability region, $\Omega_m$ of the intermediate-speed control loop. At this stage, the fastest loop is reduced to an algebraic equation, while the slowest loop variables remain constant.

    \item Determine the stability region, $\Omega_s$ of the slowest control loop. At this stage, the other two control loops are considered fast and reduced to algebraic equations.

    \item The overall stability region of the full-order ODE system is the intersection of the three individual stability regions, i.e.,
    \[
    \Omega_{\mathrm{overall}} = \Omega_f \cap \Omega_m \cap \Omega_s.
    \]
    \item Repeat the above steps for all possible bandwidth separation choices.
\end{enumerate}
For each loop's analysis, the reduced-order DAE is of order no more than two, which simplifies the derivation of stability regions and provides physical insight, and it is the insights in this step combined with the establishment of the overall stability region accounting for interaction which are the advantages of this approach.

\section{Control Loops Analysis based on Proposed Method}
The analysis begins, in subsections A and B, considering only the PLL and DVC control loops, without the dynamics of TVC such that $i_q$ is treated as a constant (typically $i_q=0$). The effect of TVC is then introduced in subsection C.
\subsection{Bandwidth sequence: \texorpdfstring{$\omega_{\text{pll}} \text{ sufficiently larger than } \omega_{\text{dvc}}$}{}}
\subsubsection{Stability analysis}~\\
\begin{figure}[t!]
\centering
\includegraphics[width=3.2in]{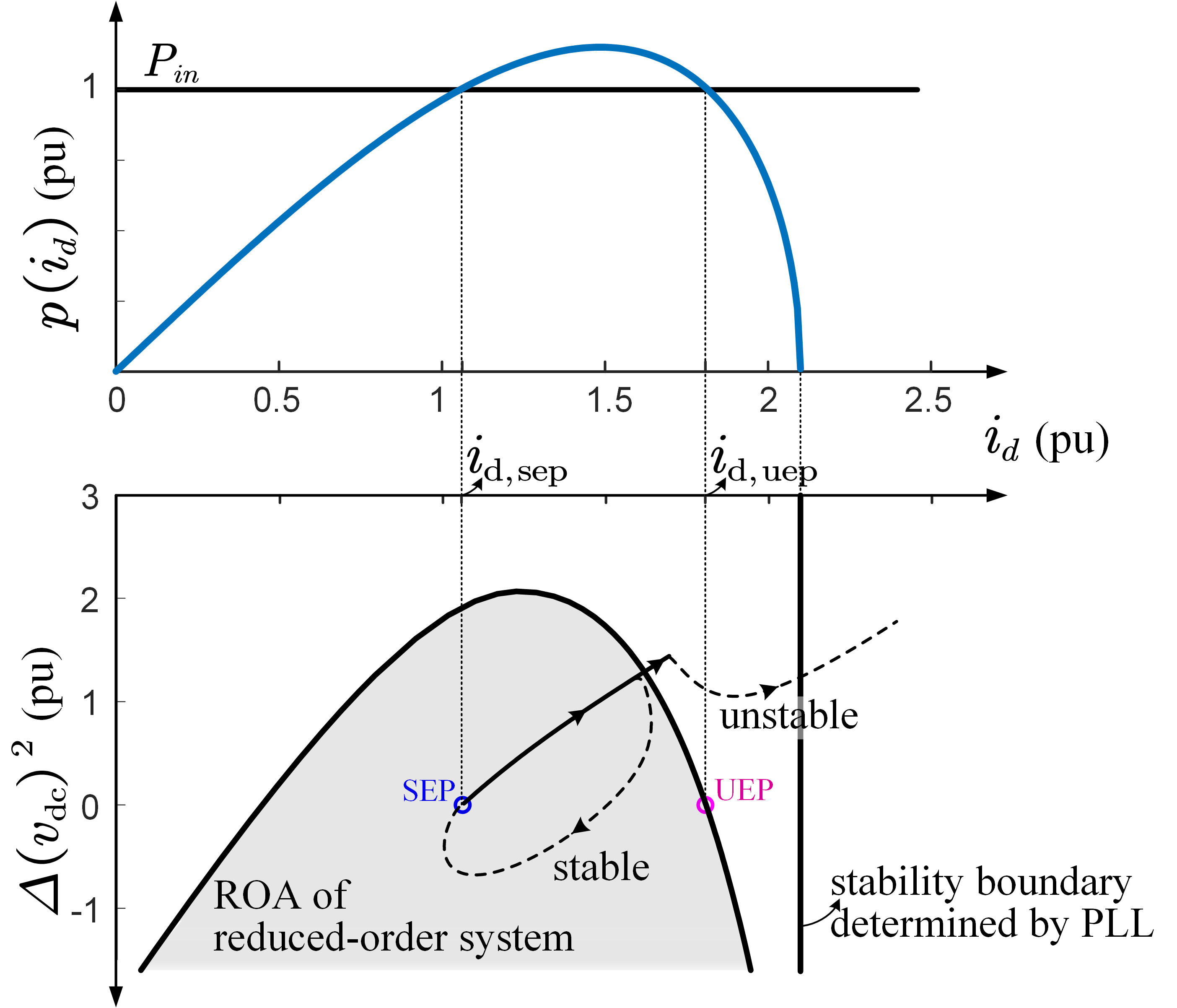}
\caption{Example curve of $p(i_d)$ (solid line on upper axes) and stability region (ROA) of DVC (lower axes) when $\omega_{\mathrm{pll}}\gg\omega_{\mathrm{dvc}}$.}
\label{DVCslow}
\end{figure} 
\indent With \texorpdfstring{$\omega_{\text{pll}} \text{ sufficiently larger than } \omega_{\text{dvc}}$}{}, the PLL subsystem \eqref{PLL} is considered as a fast subsystem, and rapidly converges to the algebraic equation \eqref{PLLfast_A}. The DVC subsystem \eqref{DVC} is treated as the slow subsystem. Substituting \eqref{PLLfast_A} into \eqref{DVC} yields a reduced-order model for the DVC dynamics, expressed in \eqref{DVCs_D} and rearranged as \eqref{PLLfast_D}. This reduced model accurately captures the system behavior, and approximates the original full-order system with an error of order $O(\varepsilon)$. The full derivation is given in Appendix B.
\begin{subequations}
\begin{equation}
    \delta = \mathrm{asin}(\frac{X_gi_d}{U_g})
    \label{PLLfast_A}
\end{equation}    
\begin{equation}
\begin{cases}
	\frac{d}{dt}i_d=k_{\mathrm{p},\mathrm{dvc}}\frac{2}{C_{\mathrm{dc}}/\omega _s}\left[ P_{in}-p\left( i_d \right) \right] +k_{\mathrm{i},\mathrm{dvc}}\Delta \left( v_{\mathrm{dc}} \right) ^2\\
	\frac{d}{dt}\Delta \left( v_{\mathrm{dc}} \right) ^2=\frac{2}{C_{\mathrm{dc}}/\omega _s}\left[ P_{in}-p\left( i_d \right) \right]\\
\end{cases}
\label{PLLfast_D}
\end{equation} 
\label{DVCslow_DAE}
\end{subequations}
where,
\begin{align*}
    p\left( i_d \right) =i_d\sqrt{{U_g}^2-\left( i_dX_g \right) ^2}-i_di_qX_g
\end{align*}

It can be shown that $p\left( i_d \right)$ is a convex function, which typically intersects the input power $P_{\mathrm{in}}$ at two points: one is a stable equilibrium point (SEP), denoted $i_{\mathrm{d,sep}}$, and the other is an unstable equilibrium point (UEP), denoted $i_{\mathrm{d,uep}}$. An example characteristic curve of $p\left( i_d \right)$ is illustrated with the solid line on upper axes of \figref{DVCslow}. The corresponding ROA, is shown as the shaded area on the lower axes in Fig.~\ref{DVCslow}. The ROA is bounded by the stable manifold of the UEP according to the stability region theory in \cite{chiang2015stability}.

To ensure that the PLL subsystem remains stable while acting as a fast loop, the current $i_d$ must satisfy: 
\small
\begin{align*}
    i_d<\min \left\{ \frac{U_g}{X_g},\frac{\sqrt{{U_g}^2-\left( X_gi_d \right) ^2}}{\zeta _{\mathrm{pll}}L_g/\omega _s},\frac{1}{\omega _{\mathrm{pll}}L_g} \right\} 
\end{align*}
\normalsize
as derived in \eqref{idreq}. This constraint, which ensures convergence of PLL dynamics, is also plotted as a vertical black line in Fig.~\ref{DVCslow}, which usually lies outside the DVC's stability region. This means that the stability of the PLL does not guarantee overall system stability if the DVC dynamics lie outside their stability region. This highlights that the slow loop can dominate the full-system behavior even when the fast loop is well-behaved. Furthermore, configuring $k_{\mathrm{i,pll}} = \zeta_{\mathrm{pll}} \cdot k_{\mathrm{p,pll}}$ results in an overdamped PLL, which expands the allowable range of $\delta$ from $(-\pi,\pi)$ (see \figref{PLLSR} in Appendix B). In summary, all these mean that the stability region for $i_d$ is shrunk compared to the standalone PLL.

Herein, the maximum output power of grid-following inverters is also discussed. The maximum value of output power is $p(i_d)_{\text{max}}=U_g^2/2X_g$  at $i_d=U_g/(\sqrt{2}X_g)$ when there is no $q$-axis current injection, $i_q=0$. If $P_{\text{in}} < p(i_d)_{\text{max}}$, namely $U_g<\sqrt{2P_{\mathrm{in}}X_g}$, does not satisfy, there is no equilibrium point, and the DVC will be unstable first before PLL. This observation aligns with the small-signal constraints on $X_g$ and $P_{\text{in}}$ discussed in \cite{huang2017effect}. In addition, a negative $i_q$ can lift the curve of $p(i_d)$, thereby expanding the stability region.

\subsubsection{Model validation}~\\
\begin{figure}[t!]
    \centering
    \subfloat[phase portrait.]{\includegraphics[width=0.28\textwidth]{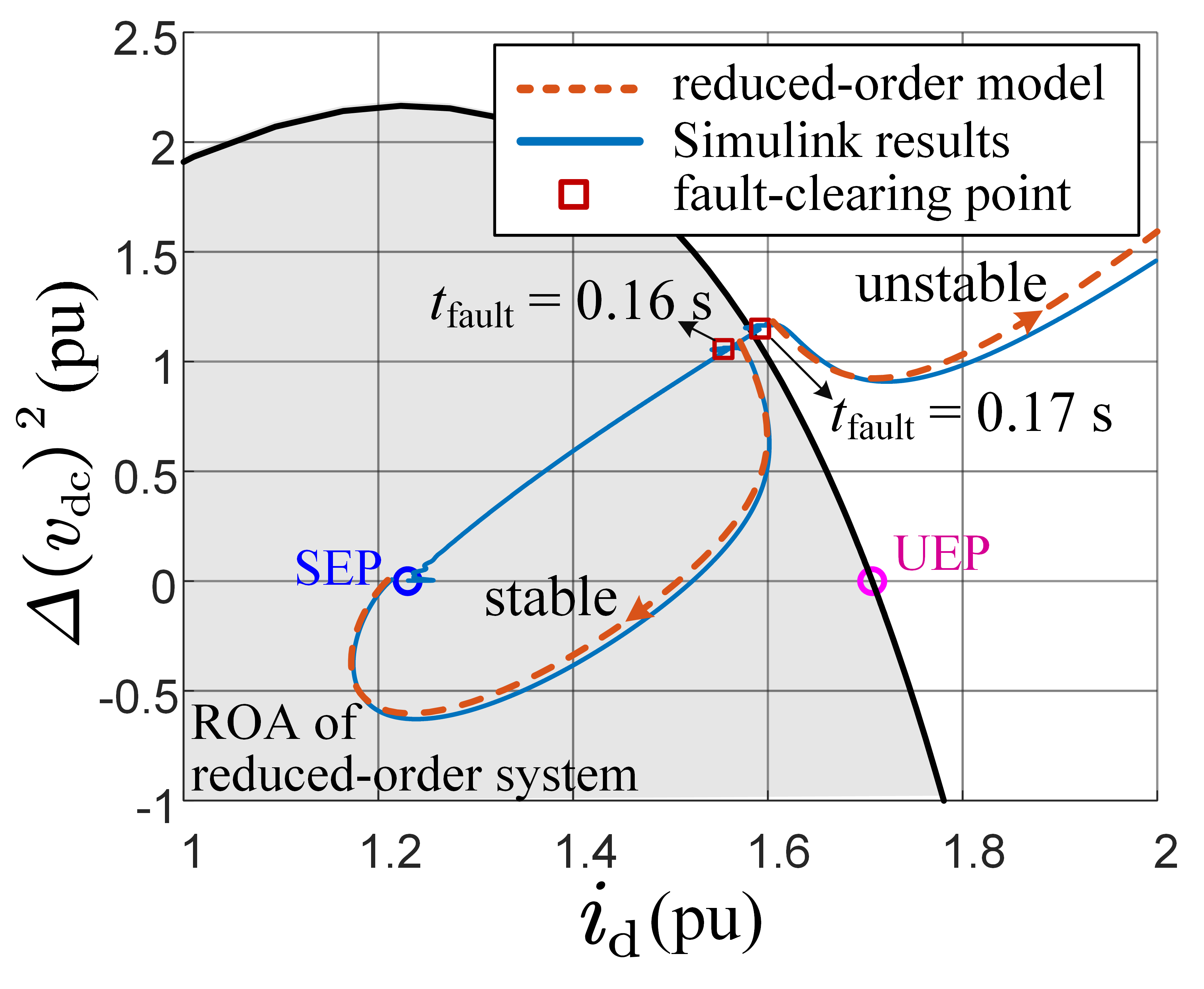}}\\
    \subfloat[Time domain comparison.]{\includegraphics[width=0.44\textwidth]{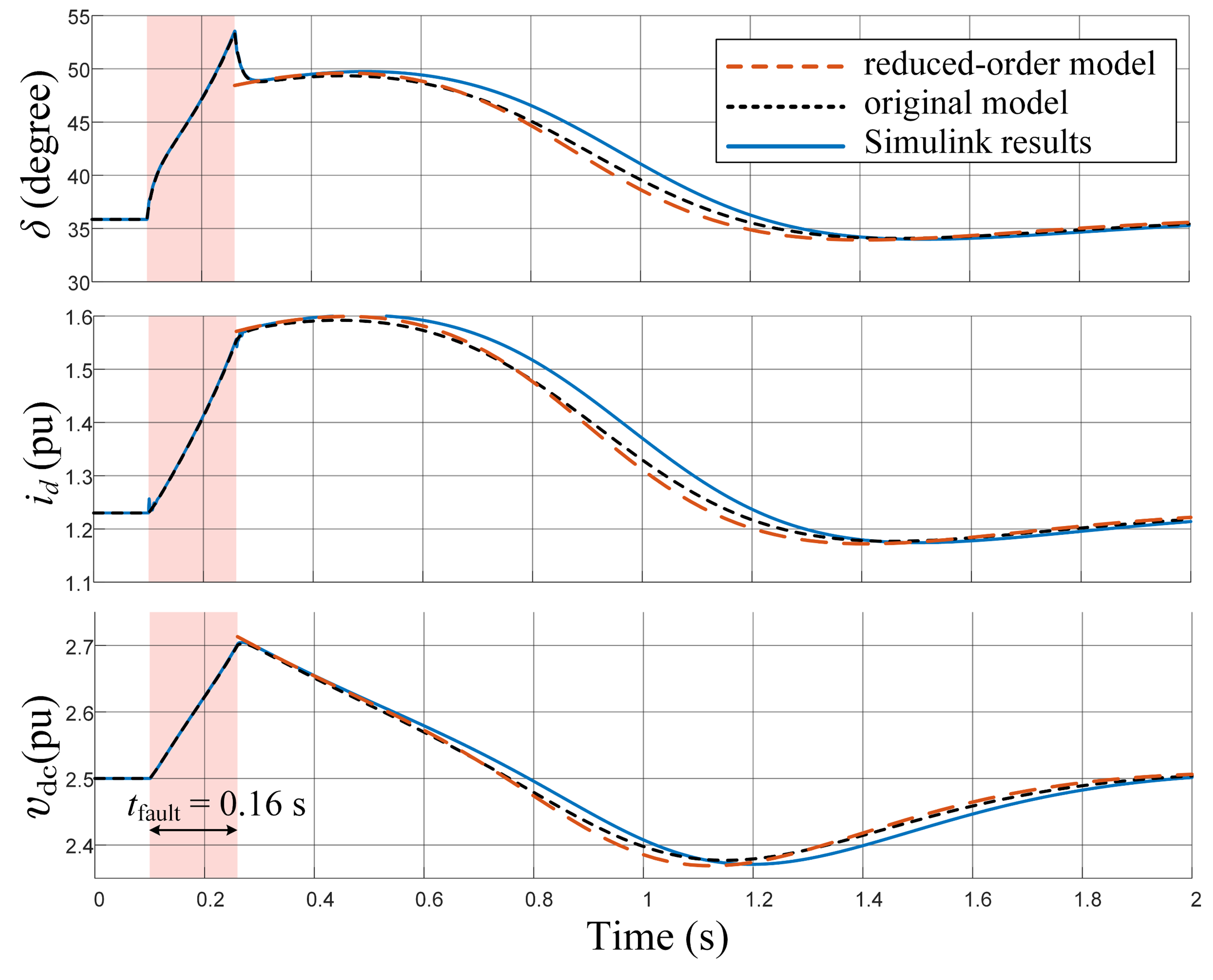}}
    \caption{Theoretical and simulation results under $\omega_{\mathrm{pll}}=15~\text{Hz}\times 2\pi$, $\omega_{\mathrm{dvc}}=2~\text{Hz}\times 2\pi$, with a 0.9~pu voltage sag fault. The corresponding CCT is approximately 0.16~s.}
    \label{Sim_DVCslow}	
\end{figure}
\indent To verify the accuracy of the model reduction, numerical results from the original ODE model \eqref{ODE}, the reduced DAE model \eqref{DVCslow_DAE}, and Simulink-based simulations are compared, which are shown in Fig.\ref{Sim_DVCslow}. Detailed simulation parameters are listed in Table~\ref{Parameter}. The PLL bandwidth is configured as 15Hz, while the DVC bandwidth is 2Hz. The critical clearing time (CCT) is found to be $t_{\text{fault}} = 0.16\text{s}$; when the clearing time increases to 0.17s, the system becomes unstable, as illustrated in Fig.\ref{Sim_DVCslow}~(a).

From the time-domain comparison in Fig.~\ref{Sim_DVCslow}(b), the reduced-order model closely matches the full-order ODE model. For the variable $\delta$, the full-order model (black dotted line) rapidly converges to the algebraic approximation \eqref{PLLfast_A} (orange dashed line), confirming the fast dynamics' behavior. For $i_d$ and $\Delta \left( v_{\mathrm{dc}} \right) ^2 = v_{\mathrm{dc}}^2 - V_{\mathrm{dc,ref}}^2$, the reduced model \eqref{PLLfast_D} achieves an $O(\varepsilon)$ approximation of the full-order model. To improve accuracy at the initial instant, a first-order correction is applied, yielding $O(\varepsilon^2)$ accuracy. Specifically, the initial values for $i_d$ and $\Delta \left( v_{\mathrm{dc}} \right) ^2$ are corrected as $i_d(0) - i_d^*(0)$ and $\Delta \left( v_{\mathrm{dc}} \right) ^2(0) - i_d^*(0)$ and $[\Delta \left( v_{\mathrm{dc}} \right) ^2]^*(0)$, respectively. The expressions for $i_d^*(0)$ and $[\Delta \left( v_{\mathrm{dc}} \right) ^2]^*(0)$ are provided by \eqref{initialvaluedvc} in Appendix~\ref{B}, accounting for the boundary-layer dynamics induced by the fast convergence of $\delta$.

The minor discrepancies between the original model and the Simulink simulation are primarily attributed to unmodeled dynamics such as the inner current control loop and the LCL filter, which are not captured in the reduced-order formulation.

\subsection{Bandwidth sequence: \texorpdfstring{$\omega_{\mathrm{dvc}} \text{ sufficiently larger than } \omega_{\mathrm{pll}}$}{}}
\subsubsection{Stability analysis}~\\
\begin{figure}[t!]
\centering
\includegraphics[width=3.2in]{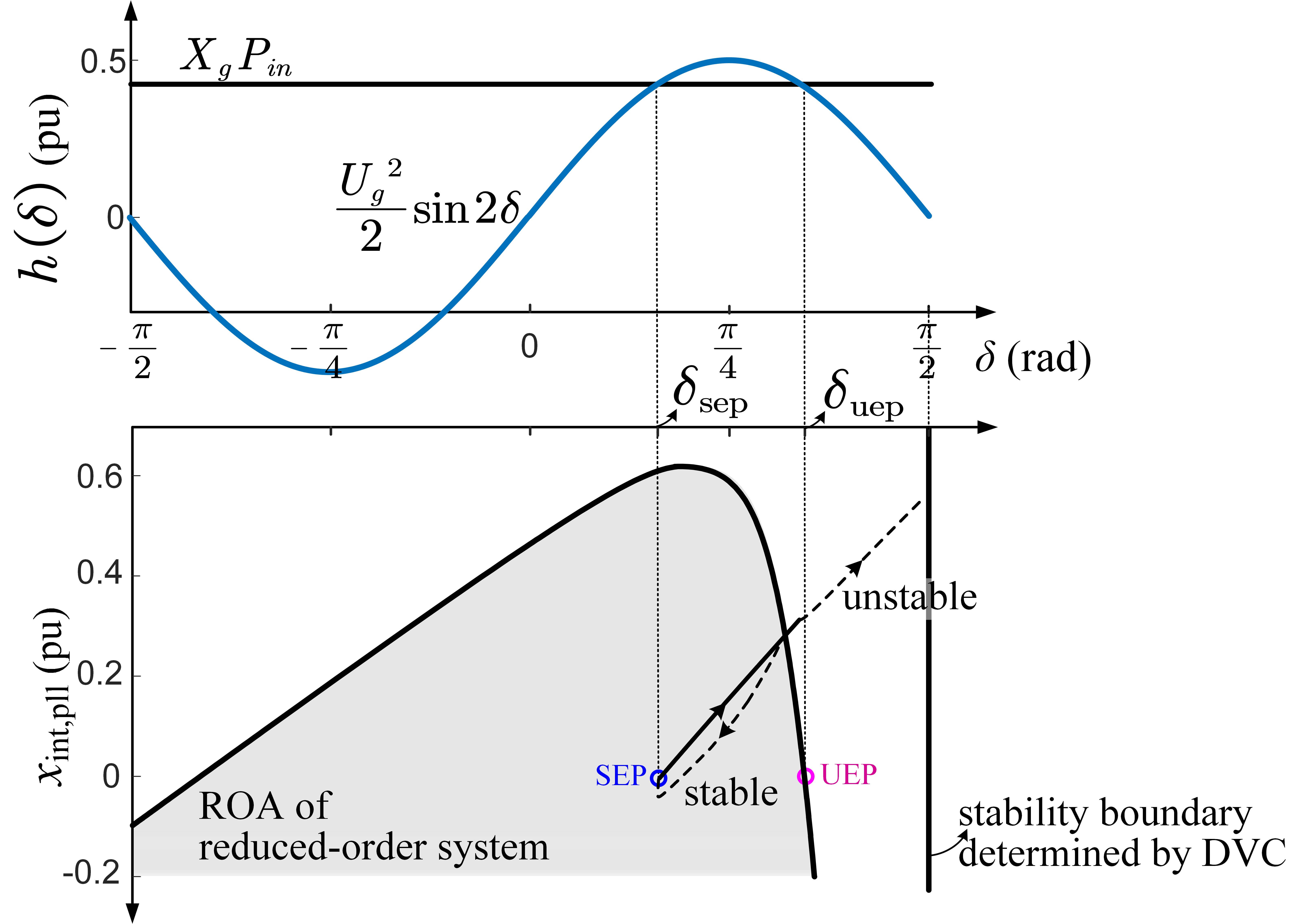}
\caption{Curve of $h(\delta)$ and stability region of PLL when $\omega_{\mathrm{dvc}}\gg\omega_{\mathrm{pll}}$.}
\label{PLLslow}
\end{figure} 
It can be shown (see Appendix~\ref{B}) that when $\delta > 0$, the DVC dynamics are always stable. The resulting algebraic equation of DVC is given by \eqref{DVCfast_A}. Substituting this into the PLL dynamics \eqref{PLL} yields a reduced-order model for the slow PLL subsystem, described in \eqref{PLLslow_DAE}. This reduced model provides an accurate approximation of the full-order system, with an error of order $O(\varepsilon)$. The full derivation and boundary-layer corrections are detailed in Appendix B.
\begin{subequations}
\begin{equation}
    i_d=\frac{P_{in}+i_qU_g\sin \delta}{U_g\cos \delta}
    \label{DVCfast_A}
\end{equation}    
\begin{equation}
\begin{cases}
	\dot{\delta}=\frac{k_{\mathrm{p},\mathrm{pll}}}{\lambda \left( \delta \right)}\cdot \left[ X_gP_{in}-h\left( \delta \right) \right] +\frac{P_{in}}{\lambda \left( \delta \right)}\cdot x_{\mathrm{int},\mathrm{pll}}\\
	\dot{x}_{\mathrm{int}\_pll}=\frac{k_{\mathrm{i},\mathrm{pll}}}{\lambda \left( \delta \right)}\cdot \left[ X_gP_{in}-h\left( \delta \right) +P_{in}L_g\cdot x_{\mathrm{int},\mathrm{pll}} \right]\\
\end{cases}
\label{DVCfast_D}
\end{equation} 
\label{PLLslow_DAE}
\end{subequations}
where,
\begin{align*}
    h\left( \delta \right) =&\frac{{U_g}^2}{2}\sin 2\delta -i_qX_gU_g\sin \delta \\
=&\frac{{U_g}^2}{2}\sin 2\delta \,\, \text{(if $i_q=0$)}
\end{align*}
\begin{align*}
\lambda \left( \delta \right) =U_g\cos \delta -k_{\mathrm{p},\mathrm{pll}}L_g/\omega _s\cdot \left( P_{in}+i_qU_g\sin \delta \right) 
\end{align*}
The function $X_g P_{\mathrm{in}}$ and the curve $h(\delta)$ are plotted in \figref{PLLslow}. The two intersection points, $\delta_{\mathrm{sep}}$ and $\delta_{\mathrm{uep}}$, correspond to the SEP and UEP, respectively. It can be shown that $\lambda(\delta)$ remains positive due to the limitation for current and PLL bandwidth. Compared to the standalone PLL system, the sinusoidal function is reduced to a period of $\pi$, resulting in $\delta_{\mathrm{sep}} \in (0, \pi/4)$ and $\delta_{\mathrm{uep}} \in (\pi/4, \pi/2)$. The corresponding stability region of the reduced-order system, bounded by the stable manifold of the UEP \cite{chiang2015stability}, is shown as the gray shaded area in Fig.~\ref{PLLslow}. Relative to the standalone PLL case, this stability region is visibly reduced. Additionally, the feasible region defined by the DVC stability condition $\cos\delta > 0$ is indicated by the vertical black line. Typically, this constraint lies outside the reduced-order system’s stability region.
 
\subsubsection{Model validation}~\\
\begin{figure}[t!]
    \centering
    \subfloat[Phase portrait.]{\includegraphics[width=0.28\textwidth]{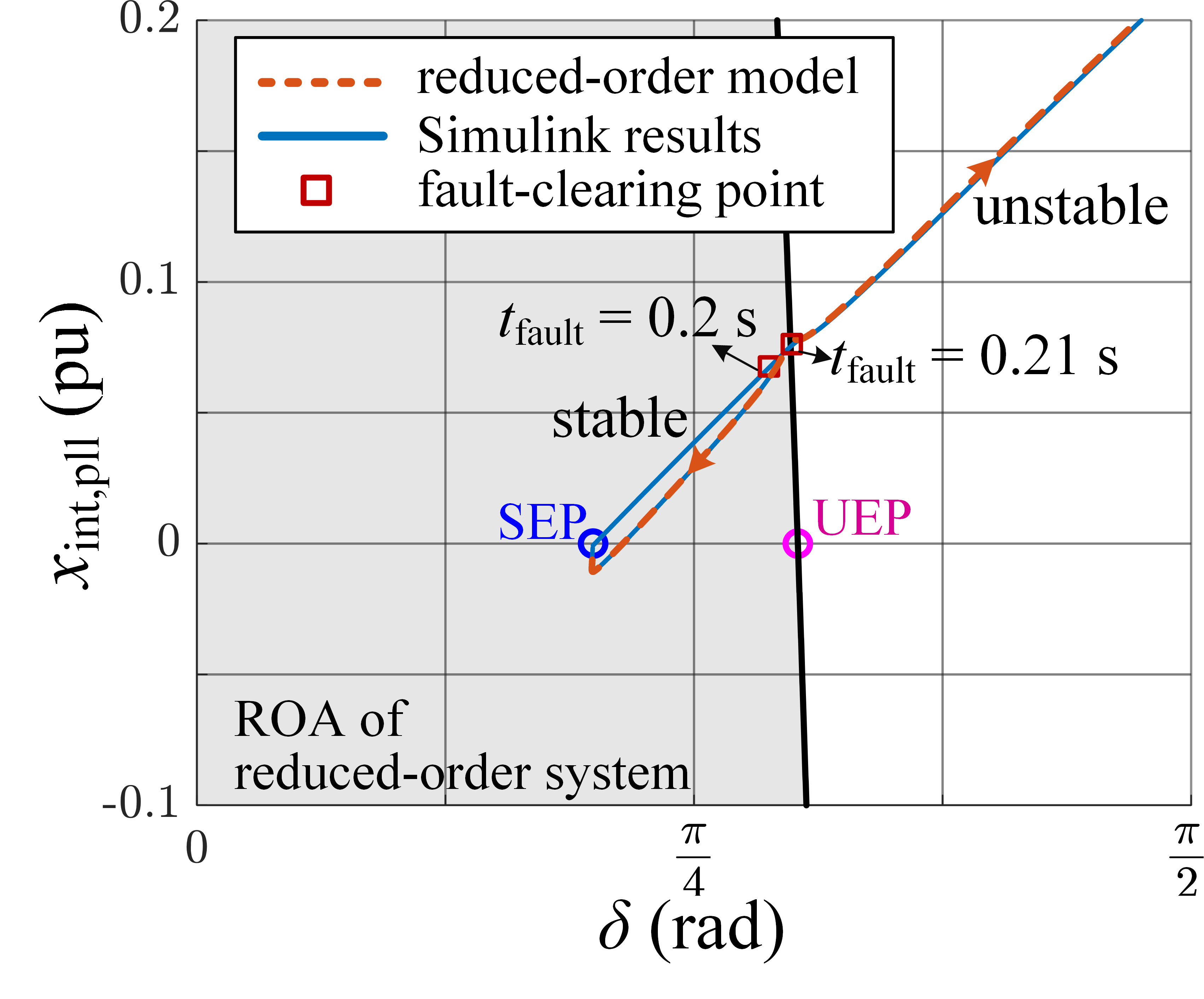}}\\
    \subfloat[Time domain comparison.]{\includegraphics[width=0.44\textwidth]{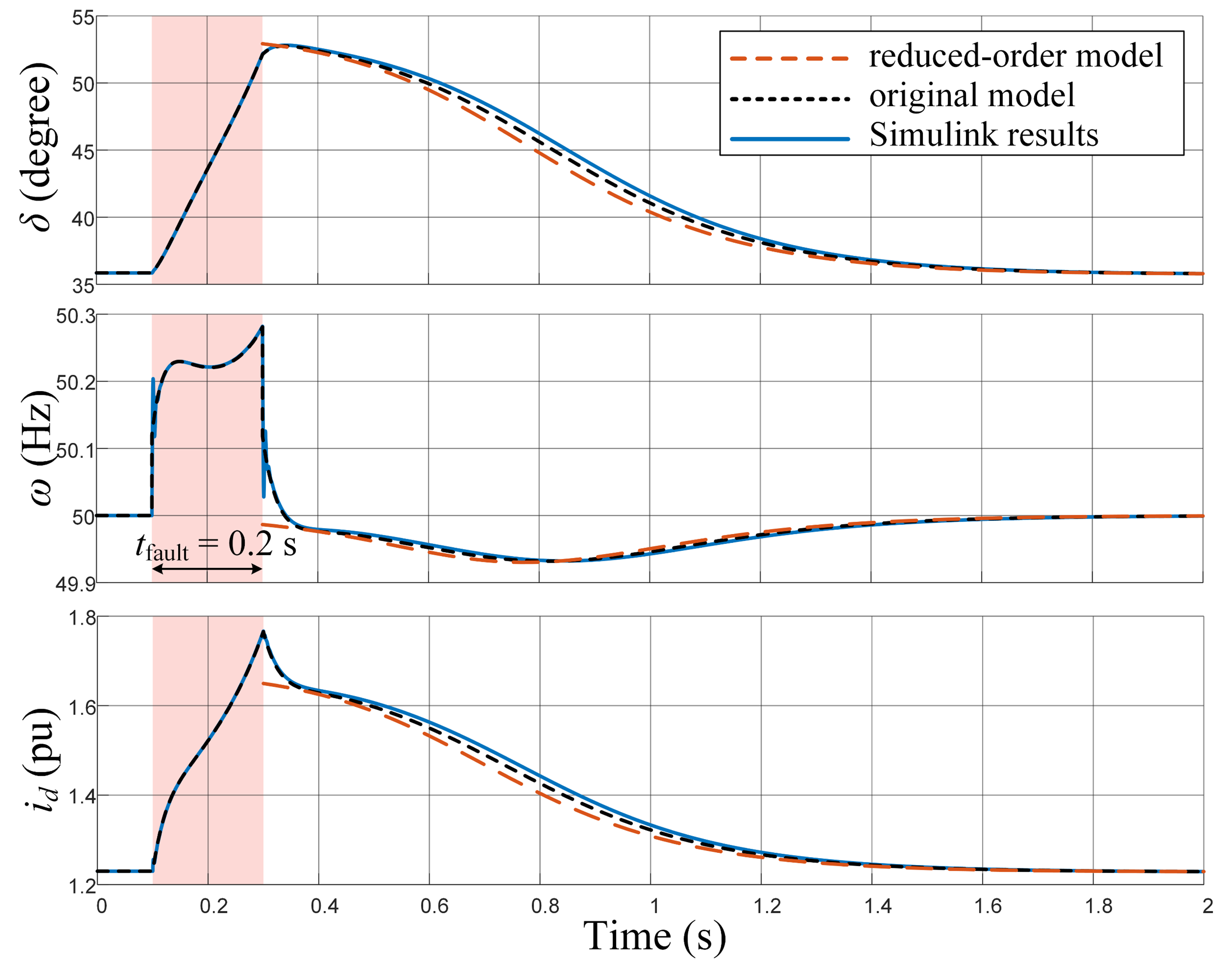}}
    \caption{Theoretical and simulation results under $\omega_{\mathrm{pll}}=2~\text{Hz}\times 2\pi$, $\omega_{\mathrm{dvc}}=15~\text{Hz}\times 2\pi$, with a 0.9~pu voltage sag fault. The corresponding CCT is approximately 0.2~s.}
    \label{Sim_PLLslow}	
\end{figure}
\indent To verify the accuracy of the model reduction, numerical results from the original ODE model \eqref{ODE}, the reduced DAE model \eqref{PLLslow_DAE}, and Simulink-based simulations are compared, which are shown in Fig.\ref{Sim_DVCslow}. Detailed simulation parameters are listed in Table~\ref{Parameter}. From the time-domain comparison in Fig.~\ref{Sim_PLLslow}(b), the reduced-order model closely matches the full-order ODE model. For the variable $i_d$, the full-order model (black dotted line) rapidly converges to the algebraic approximation \eqref{DVCfast_A} (orange dashed line), confirming the fast dynamics behavior of the DVC loop. To improve accuracy at the initial instant, a first-order correction is applied, yielding $O(\varepsilon^2)$ accuracy. Specifically, the initial values for $\delta$ and $x_{\text{int,pll}}$ are corrected as $\delta(0) - \delta^*(0)$ and $x_{\text{int,pll}}(0) -x_{\text{int,pll}}^*(0)$, respectively. The expressions for $\delta^*$ and $x_{\text{int,pll}}^*$ are provided by \eqref{initialvaluedvc} in Appendix B, accounting for the boundary-layer dynamics induced by the fast convergence of $i_d$.

\begin{table*}[hbp]
\centering
\footnotesize
\captionsetup{font=small}
\caption{ROA Comparison and Parameters Feasible Region under Different Control Loop Configurations}
\label{table1}
\renewcommand{\arraystretch}{2.2}
\begin{tabular}{c|c|c|c|c|c|c}
\hline\hline
\multirow{2}{*}{Parameter} & \multicolumn{2}{c|}{\multirow{2}{*}{\makecell{PLL-alone \\ ($i_d$ is fixed)}}} & \multicolumn{2}{c|}{\makecell{Bandwidth sequence $\omega_{\text{pll}}\gg\omega_{\text{dvc}}$}} & \multicolumn{2}{c}{\makecell{Bandwidth sequence $\omega_{\text{dvc}}\gg\omega_{\text{pll}}$}} \\
\cline{4-7}
 & \multicolumn{2}{c|}{} & No / slow TVC & Fast TVC  & No / slow TVC & Fast TVC \\
\hline
Grid voltage $U_g$ & \multicolumn{2}{c|}{$U_g > i_dX_g$} & \makecell{ $U_g>\sqrt{2P_{\text{in}}X_g}$ \\ {[worse]}} & [mitigated] & \makecell{ $U_g>\sqrt{2P_{\text{in}}X_g}$ \\ {[worse]}} & [mitigated] \\
\hline
ROA of DVC state & \multicolumn{2}{c|}{$i_d<i_{\text{d,max}}$} & [reduced] & [mitigated] & \makecell{whole DVC\\ phase plane} & \makecell{whole DVC\\ phase plane} \\
\hline
ROA of PLL state & \multicolumn{2}{c|}{ $\delta < \delta_{\text{uep,pll}}$} & \makecell{same as PLL-alone}  & \makecell{same as PLL-alone}  & [reduced] & [mitigated] \\
\hline\hline
\end{tabular}
\end{table*}

\subsection{Impact of TVC control loop}
According to \eqref{ODE}, the TVC loop has no direct influence on the PLL, as the PLL dynamics are independent of $i_q$. Therefore, the TVC primarily affects the DVC subsystem. Consequently, the bandwidth relationship between the DVC and TVC loops becomes critical. If the TVC bandwidth is lower than that of the DVC, $i_q$ can be regarded as quasi-static from the perspective of the DVC dynamics and thus has negligible impact on the analysis presented in the previous two sections. However, when the TVC bandwidth exceeds that of the DVC, there are two possible scenarios:
\subsubsection{Bandwidth sequence: \texorpdfstring{$\omega_{\mathrm{tvc}} \text{ and } \omega_{\mathrm{pll}} \gg \omega_{\mathrm{dvc}}$}{}}~\\
\indent As shown in Appendix B, regardless of whether the PLL bandwidth is higher or lower than that of the TVC, the TVC subsystem always remains stable. As a result, the PLL and TVC loops can be jointly treated as a fast subsystem and represented by a set of algebraic equations \eqref{PLLfastTVC_A1} and \eqref{PLLfastTVC_A2}. By substituting these algebraic expressions into the original DVC model \eqref{DVC}, a new reduced-order differential equation for the DVC dynamics is obtained, as expressed in \eqref{PLLfastTVC_D}.
\begin{subequations}
\begin{equation}
    \delta =\mathrm{asin} \left( \frac{X_gi_d}{U_g} \right) 
    \label{PLLfastTVC_A1}
\end{equation}    
\begin{equation}
    -i_q=\frac{k_v}{k_vX_g+1}\cdot \left( V_{ref}-U_g\cos \delta \right) 
    \label{PLLfastTVC_A2}
\end{equation}  
\begin{equation}
\begin{cases}
	\frac{d}{dt}i_d=k_{\mathrm{p},\mathrm{dvc}}\frac{2}{C_{\mathrm{dc}}/\omega _s}\left[ P_{in}-p^{\prime}\left( i_d \right) \right] +k_{\mathrm{i},\mathrm{dvc}}\varDelta \left( v_{\mathrm{dc}} \right) ^2\\
	\frac{d}{dt}\varDelta \left( v_{\mathrm{dc}} \right) ^2=\frac{2}{C_{\mathrm{dc}}/\omega _s}\left[ P_{in}-p^{\prime}\left( i_d \right) \right]\\
\end{cases}
\label{PLLfastTVC_D}
\end{equation} 
\label{PLLfastTVC_DAE}
\end{subequations}
where,
\begin{align*}
    p^\prime&=\underset{\mathrm{synchronous}\ \mathrm{power}}{\underbrace{\frac{U_gV_{ref}}{X_g+1/k_v}\sin \delta }}+\underset{\mathrm{saliency}\ \mathrm{power}}{\underbrace{\left( \frac{1}{X_g}-\frac{1}{X_g+1/k_v} \right) \frac{{U_g}^2}{2}\sin 2\delta }}\\
	=&\frac{1}{k_vX_g+1}\cdot \underset{p\left( i_d \right)}{\underbrace{i_d\sqrt{U_g-\left( i_dX_g \right) ^2}}}+\frac{k_vX_g}{k_vX_g+1}\cdot V_{ref}i_d\\
\end{align*}

\begin{figure}[t!]
\centering
\includegraphics[width=3in]{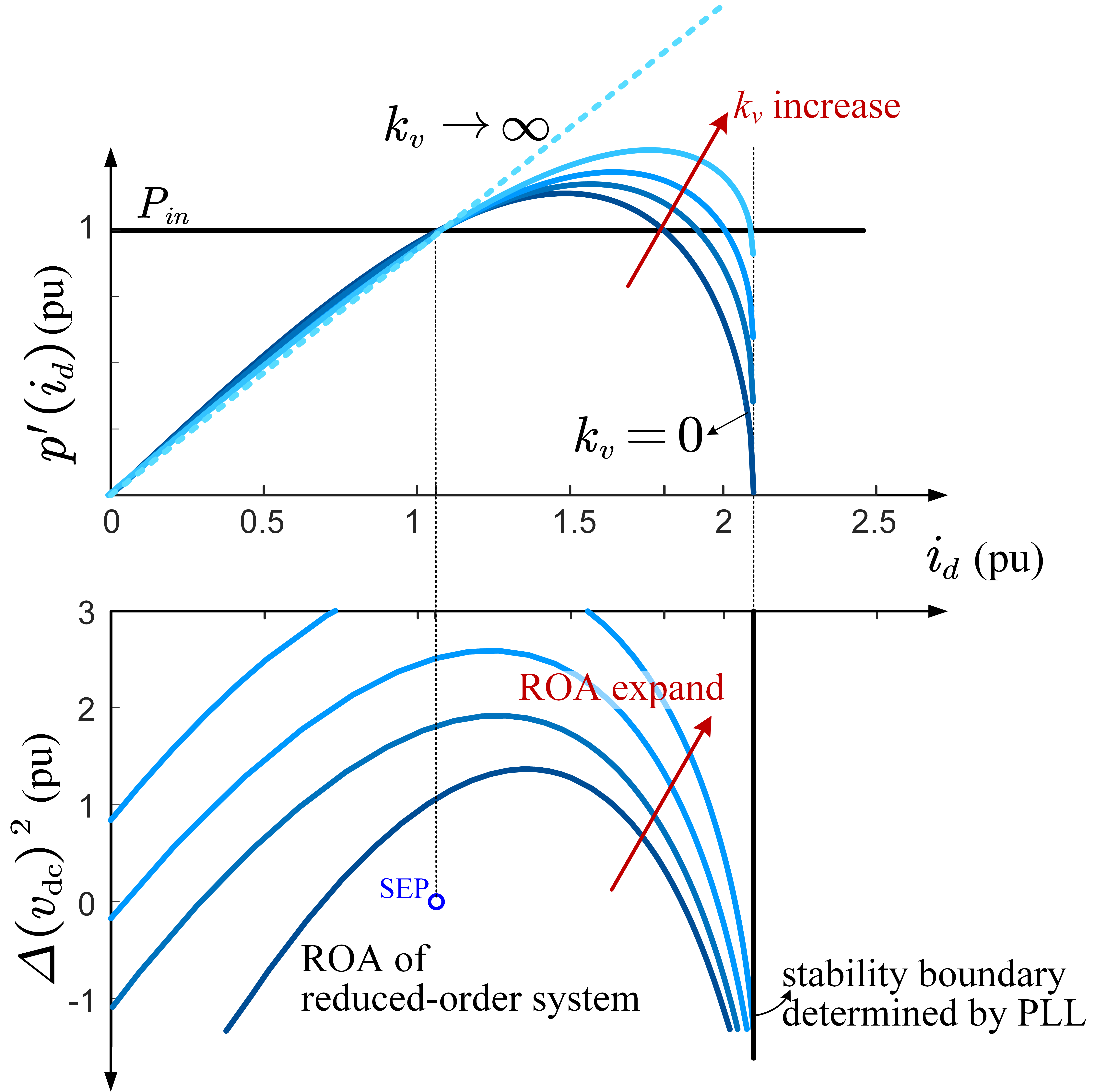}
\caption{Curve of $p^\prime(i_d)$ and stability region of DVC when $\omega_{\mathrm{tvc}} \text{ and } \omega_{\mathrm{pll}} \gg \omega_{\mathrm{dvc}}$.}
\label{DVCslowtvc}
\end{figure}

Compared to \eqref{DVCslow_DAE}, the power function $p(i_d)$ in \eqref{PLLfastTVC_DAE} is modified to $p^\prime(i_d)$, as shown in \figref{DVCslowtvc}. The curve of $p^\prime(i_d)$ is elevated relative to $p(i_d)$, without altering the steady-state operating point of the system. As a result, the stability region of the reduced-order system is expanded, as illustrated in \figref{DVCslowtvc}. The expression of $p^\prime(i_d)$ structurally resembles the power expression of a salient-pole synchronous generator \cite{kundur1994power}. In this analogy, $1/k_v$ can be interpreted as an additional impedance on the $d$-axis, and $V_{\text{ref}}$ acts as the internal voltage reference. As $k_v\rightarrow \infty$, the terminal voltage is well regulated to $V_{\text{ref}}$, and the output power $p^\prime$ approaches a straight line through the SEP: $i_d\cdot V_{\text{ref}}$. In this sense, the DVC subsystem becomes decoupled from the PLL and remains always stable. Consequently, the overall system is stable if the fast PLL subsystem is stable.

\subsubsection{Bandwidth sequence: \texorpdfstring{$\omega_{\mathrm{tvc}} \gg \omega_{\mathrm{dvc}} \gg \omega_{\mathrm{pll}}$}{}}~\\
\begin{figure}[t!]
\centering
\includegraphics[width=3in]{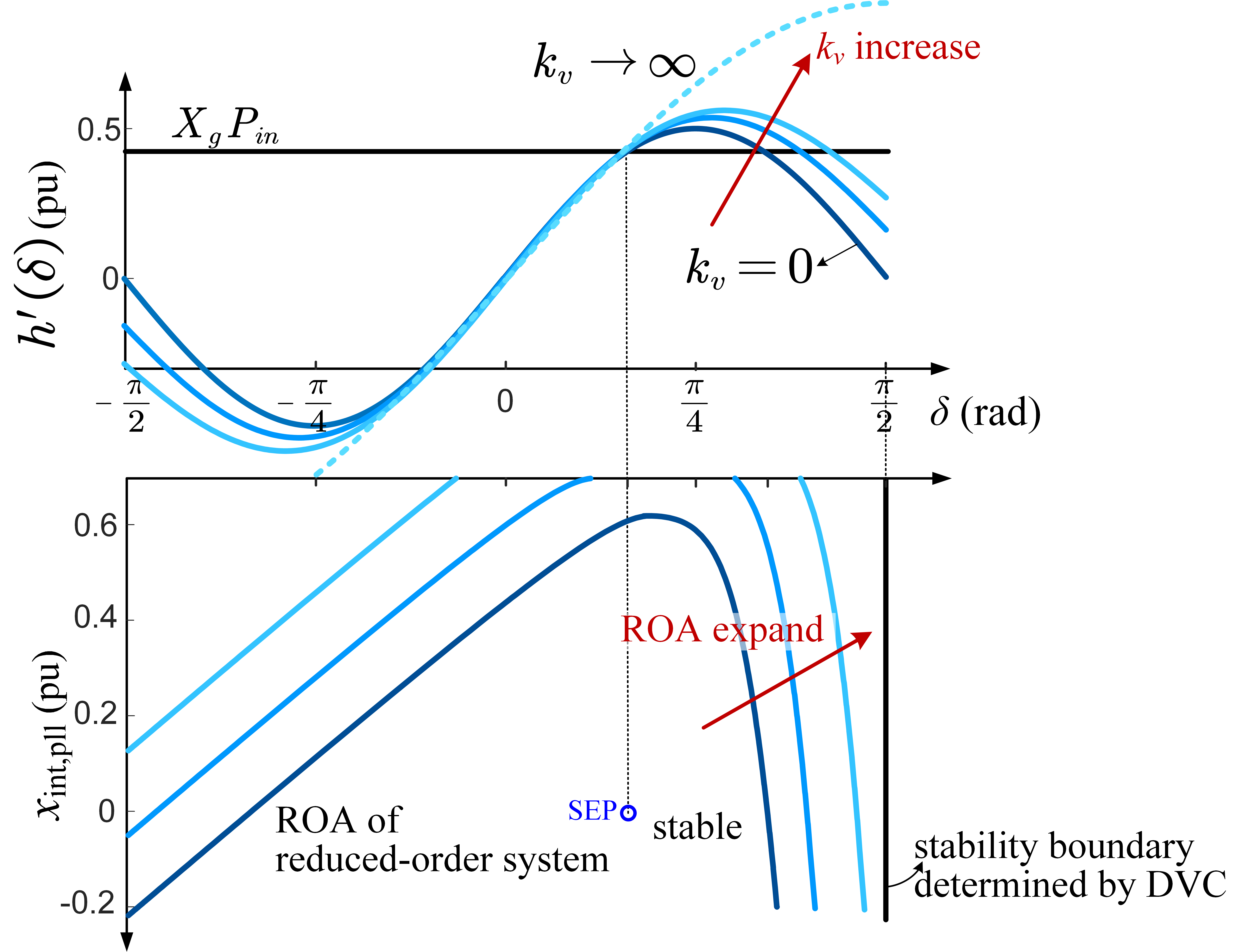}
\caption{Curve of $h^\prime(\delta)$ and stability region of PLL when $\omega_{\mathrm{tvc}} \gg \omega_{\mathrm{dvc}} \gg \omega_{\mathrm{pll}}$.}
\label{PLLslowtvc}
\end{figure}

\begin{figure}[t!]
\centering
\subfloat[$\omega_{\text{dvc}}=2~\text{Hz}\times2\pi$, \\$\omega_{\text{pll}}=15~\text{Hz}\times2\pi$.]{\includegraphics[width=0.223\textwidth]{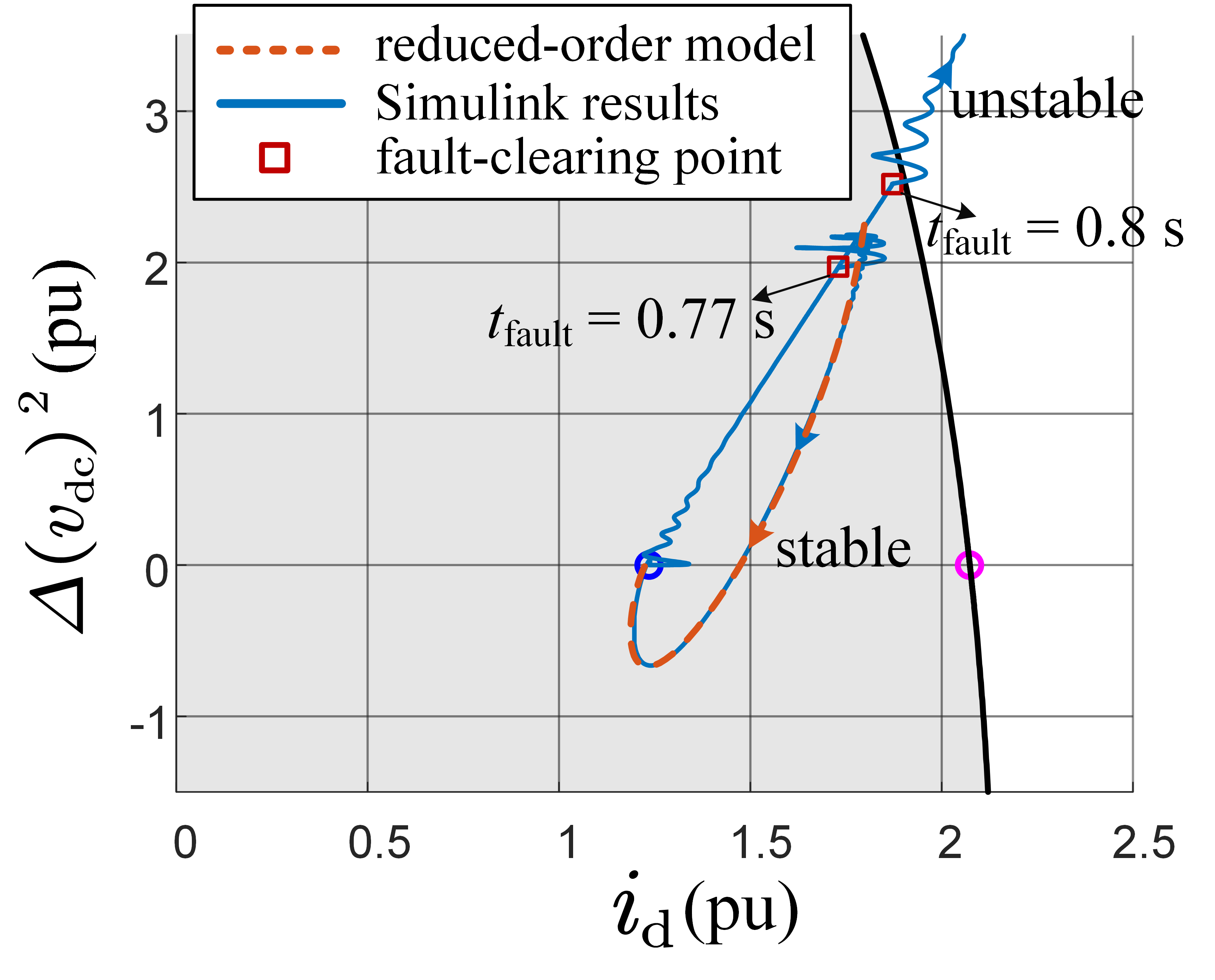}}
\subfloat[$\omega_{\text{pll}}=2~\text{Hz}\times2\pi$,\\ $\omega_{\text{dvc}}=15~\text{Hz}\times2\pi$.]{\includegraphics[width=0.245\textwidth]{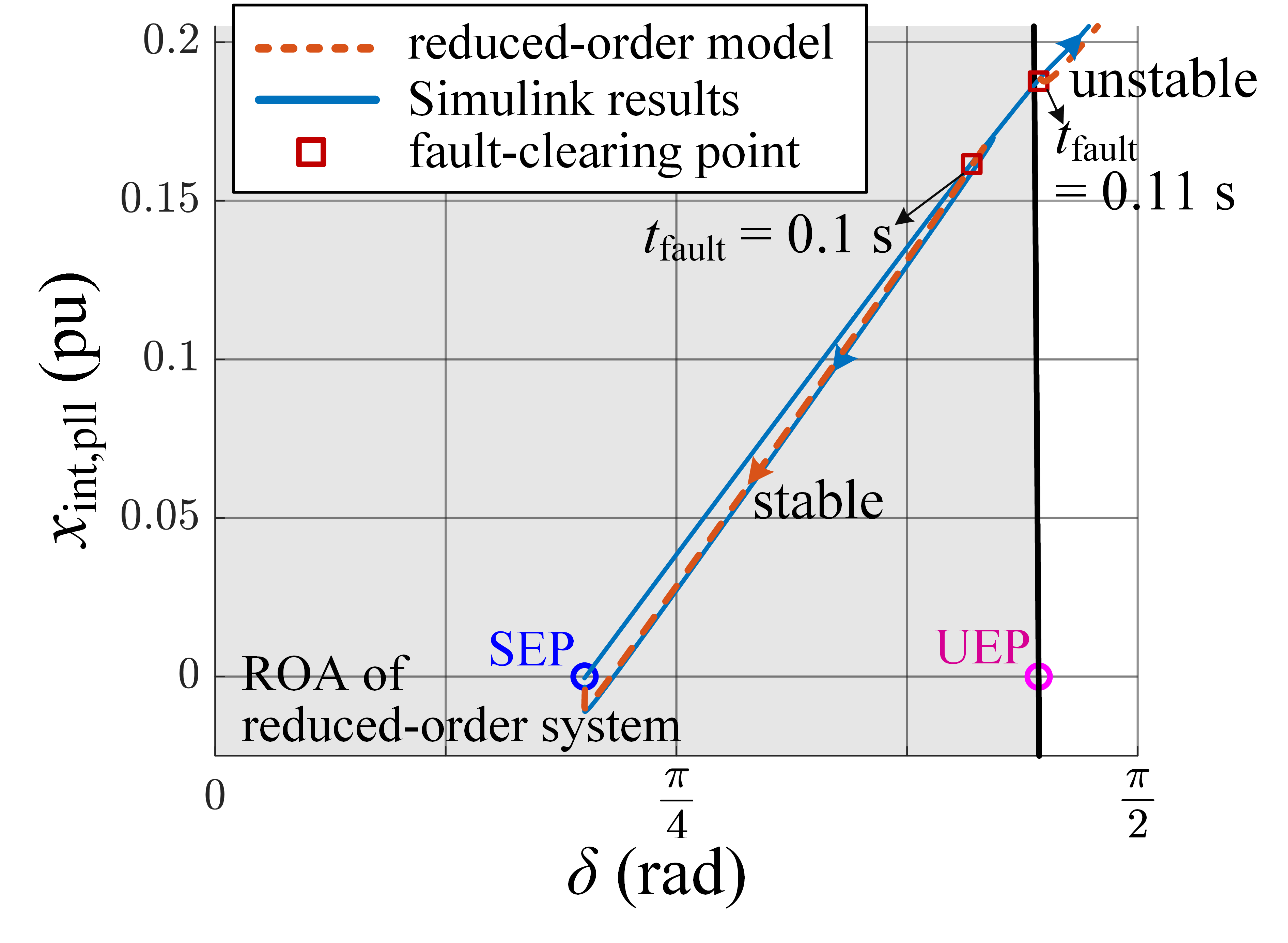}}
\caption{The phase portrait comparison between simulation and reduced-order model, with a 0.6 pu voltage sag fault, $\omega_{\text{tvc}}=20~\text{Hz}\times2\pi$, $k_v = 2$.}
\label{Sim_tvc}
\end{figure}
\indent Under this bandwidth configuration, both the DVC and TVC loops can be approximated by algebraic equations, as given in \eqref{DVCfastTVC_A1} and \eqref{DVCfastTVC_A2}. Substituting these expressions into the original PLL dynamics yields the reduced-order differential equation in \eqref{DVCfastTVC_D}.
\begin{subequations}
\begin{equation}
    i_d=\frac{P_{in}+i_qU_g\sin \delta}{U_g\cos \delta}
    \label{DVCfastTVC_A1}
\end{equation}   
\begin{equation}
    -i_q=\frac{k_v}{k_vX_g+1}\cdot \left( V_{ref}-U_g\cos \delta \right) 
    \label{DVCfastTVC_A2}
\end{equation} 
\begin{equation}
\begin{cases}
	\dot{\delta}=\frac{k_{\mathrm{p},\mathrm{pll}}}{\lambda \left( \delta \right)}\cdot \left[ X_gP_{in}-h^\prime \left( \delta \right) \right] +\frac{P_{in}}{\lambda \left( \delta \right)}\cdot x_{\mathrm{int},\mathrm{pll}}\\
	\dot{x}_{\mathrm{int}\_pll}=\frac{k_{\mathrm{i},\mathrm{pll}}}{\lambda \left( \delta \right)}\cdot \left[ X_gP_{in}-h^\prime \left( \delta \right) +P_{in}L_g\cdot x_{\mathrm{int},\mathrm{pll}} \right]\\
\end{cases}
\label{DVCfastTVC_D}
\end{equation} 
\label{DVCfastTVC_DAE}
\end{subequations}
where,
\begin{align*}
    h^\prime \left( \delta \right) =& \frac{k_vX_g}{1+k_vX_g}\cdot V_{ref}U_g\sin \delta +\frac{1}{1+k_vX_g}\cdot h\left( \delta \right) 
\end{align*}

Compared to the expression in \eqref{DVCfast_D}, the nonlinear term $h(\delta)$ is replaced by $h^\prime(\delta)$, as shown in \figref{PLLslowtvc}. The curve of $h^\prime(\delta)$ is elevated relative to $h^\prime(\delta)$, an expanded stability region for the PLL. As $k_v \rightarrow \infty$, $h^\prime(\delta)$ is the sine wave function with a period of $2\pi$ resembling the conventional PLL model, which contrasts with the original $h(\delta)$, which has a period of $\pi$.  

\subsubsection{Model validation}~\\
\indent To further validate the accuracy of the model reduction with TVC included, comparisons are made between the reduced-order models \eqref{PLLfastTVC_DAE}, \eqref{DVCfastTVC_DAE} and full Simulink simulations. The resulting trajectories are shown in Fig.\ref{Sim_tvc}. It is evident that the inclusion of the TVC loop enlarges the stability region compared to the cases without TVC. In both Fig.\ref{Sim_tvc}(a) and (b), the trajectories of the reduced-order models closely match those of the full simulation at the critical clearing times, which are approximately $t_{\text{fault}} = 0.77$~s and $0.1$~s for the two respective scenarios.

\subsection{Stability region versus bandwidth configuration}
According to the results of the analysis, the stability regions are summarized in \tabref{table1}. The stability region is shrunk compared with the PLL-alone system. The system with a high bandwidth DVC has a larger stability region on the $i_d$ dimension, while a high bandwidth PLL has a larger stability region on the $\delta$ dimension. The inclusion of a fast TVC helps alleviate the adverse effects of PLL–DVC interaction on transient stability.

\section{Analysis under Various Fault and HIL Experiment Results}
\begin{figure}[t!]
    \centering
    \subfloat[Tested system and configuration.]{\includegraphics[width=0.48\textwidth]{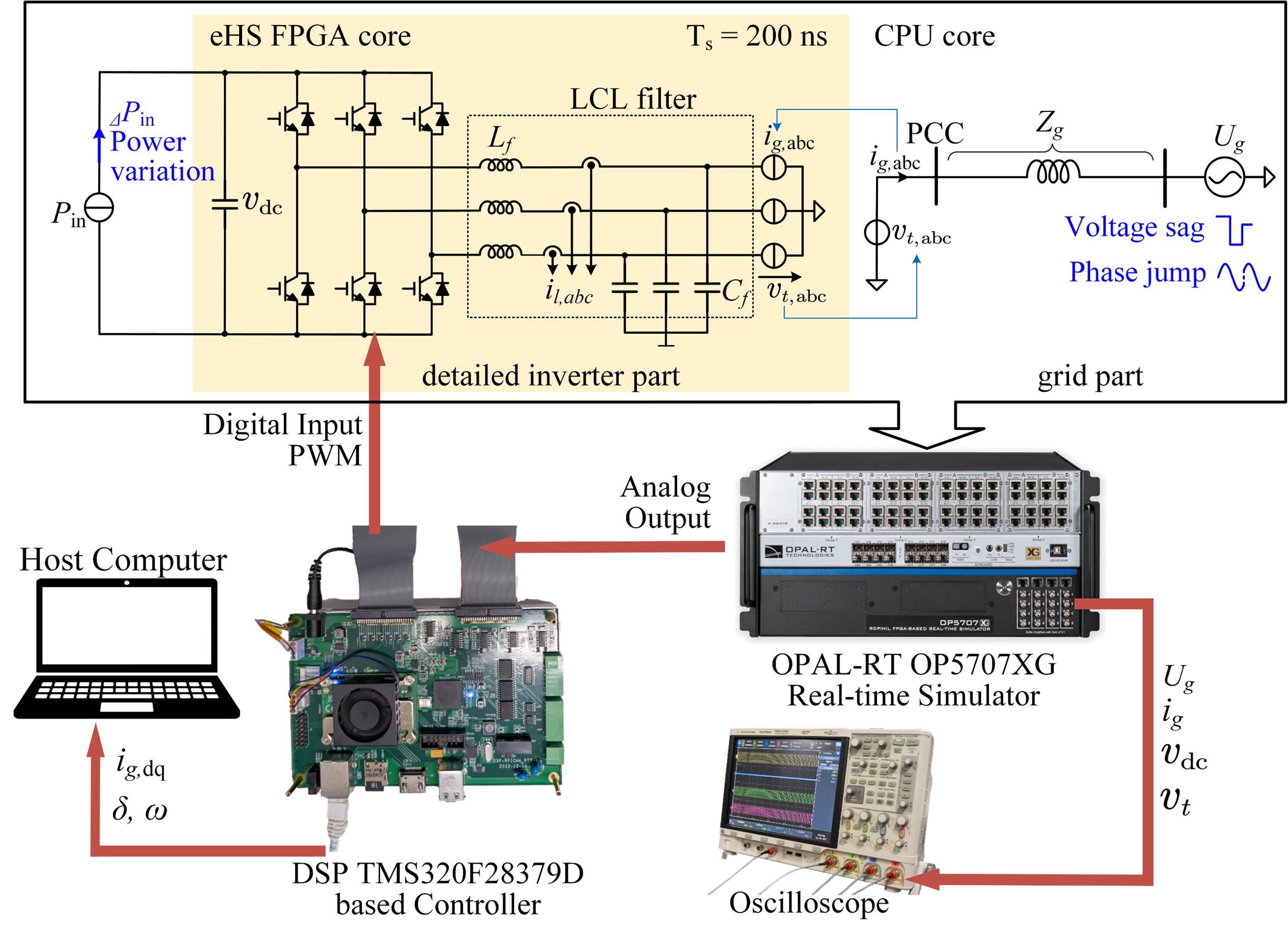}}\\
    \subfloat[Platform setup.]{\includegraphics[width=0.4\textwidth]{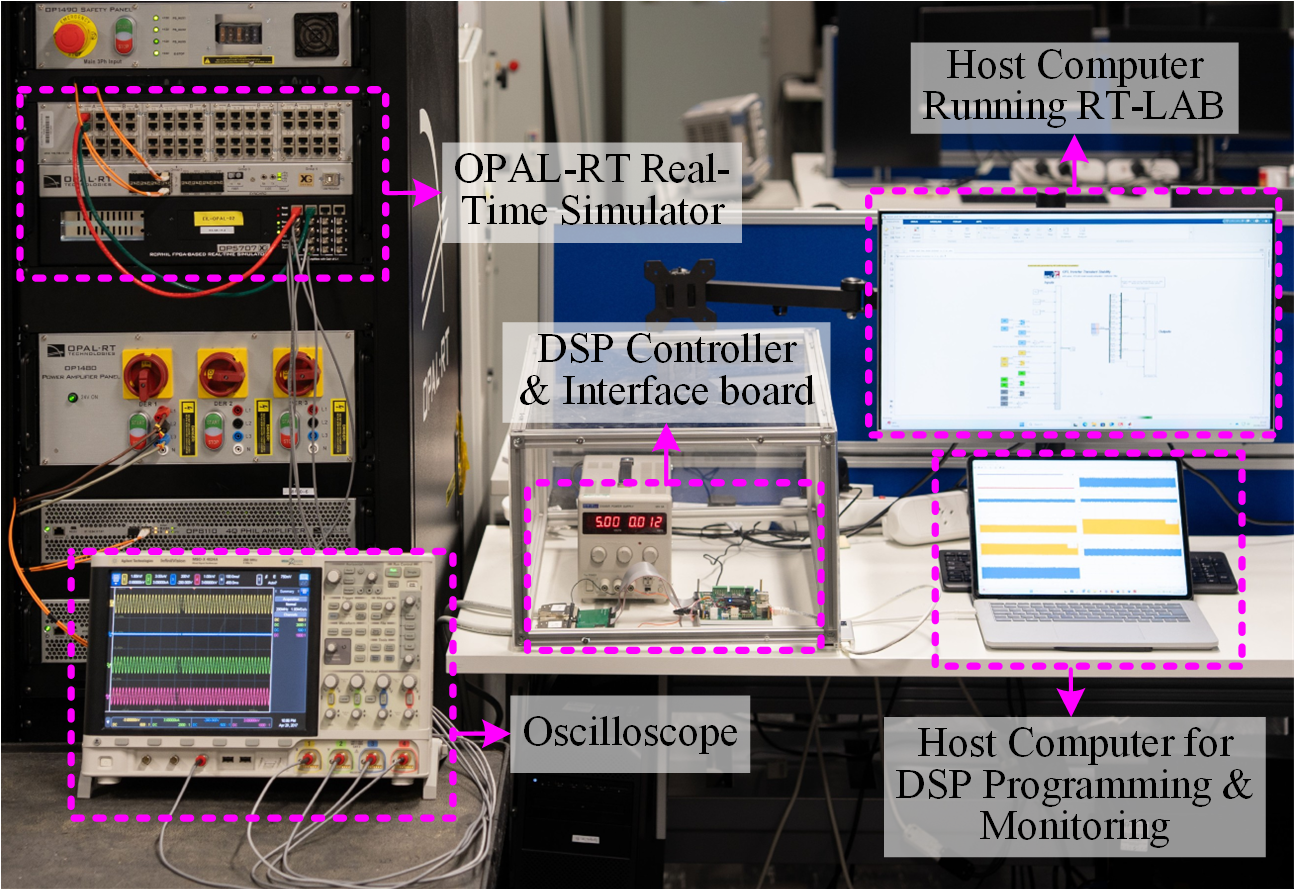}}
    \caption{HIL experimental platform.}
    \label{platform}	
\end{figure}
In order to verify the above theoretical analysis and investigate the interaction among control loops, hardware-in-the-loop (HIL) experimental results are discussed in this section. The experimental platform is shown in Fig.\ref{platform}. The setup employs the OPAL-RT real-time simulator OP5707XG equipped with the eHS FPGA-based toolbox, which can reach 200~ns real-time step, enabling highly detailed real-time modeling of the inverter. The 1~MVA GFL inverter infinite bus system is built inside the OPAL-RT real-time simulator, as depicted in \figref{platform}~(a). The detailed system parameters are listed in \tabref{Parameter} in Appendix C. Three typical faults, including voltage sag fault, phase jump, and input power variation are emulated by the OPAL-RT real-time simulator. The controller is implemented by the DSP TMS320F28379. The controller acquires analog signals from the OP5707XG and generates PWM signals to drive the IGBT switches of the GFL inverter. Both the IGBT switching frequency and the DSP control frequency are set to 10~kHz. Key signals, including grid voltage $U_{g,a}$, output current $i_{g,a}$, terminal voltage at PCC $v_{\text{t,a}}$ (all phase A), and the deviation of the DC-link voltage $v_{\text{dc}} - V_{\text{dc,ref}}$ are measured by the oscilloscope. Other control variables, such as angle difference between the PLL and the grid $\delta$, frequency of PLL $f_{\text{pll}}$, and $i_d$, $i_q$ are monitored via the host computer through DSP data upload.

\subsection{Voltage sag fault}
\begin{figure}[t!]
\centering
\includegraphics[width=0.46\textwidth]{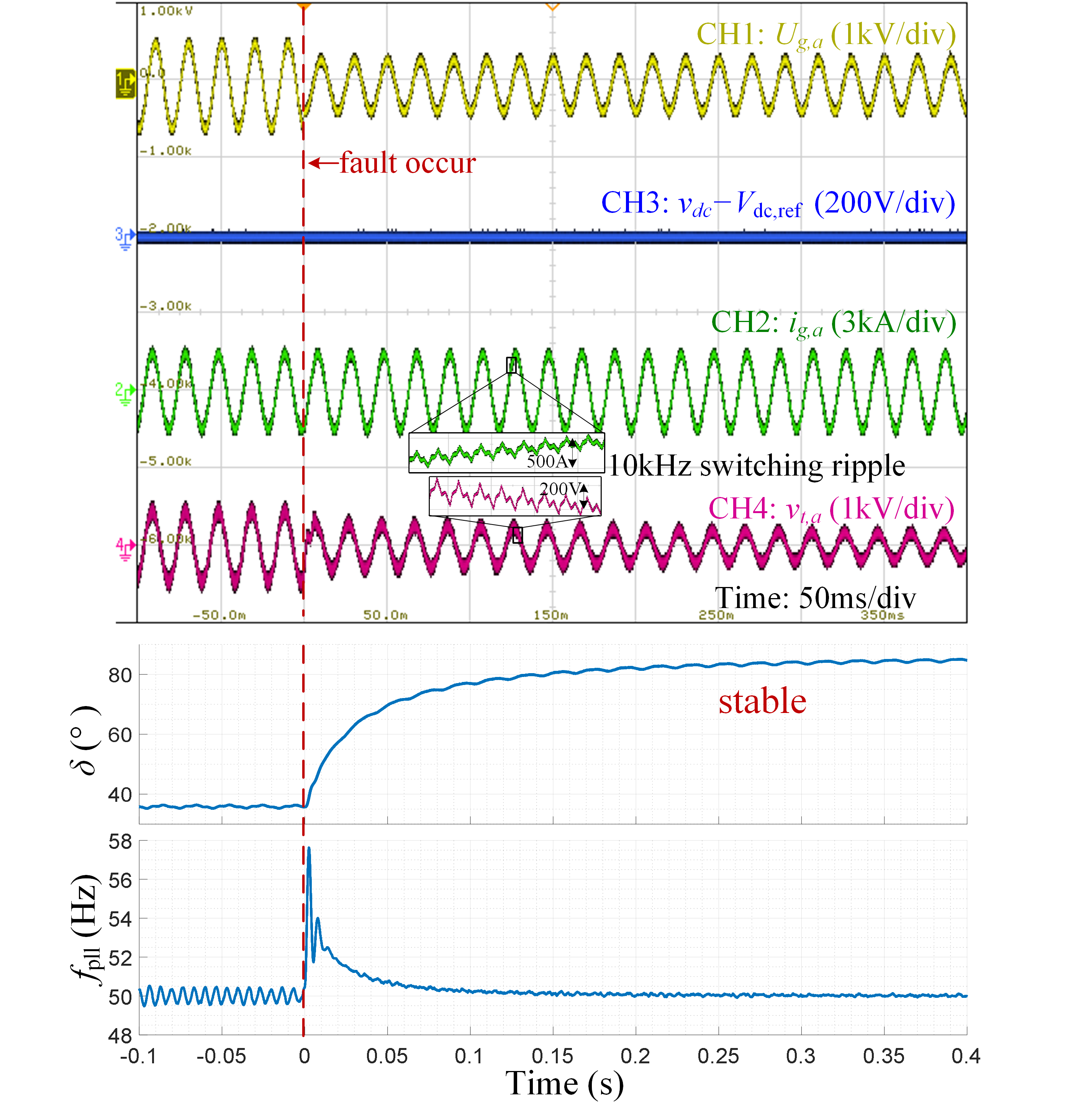}
\caption{PLL standalone system under 0.6 pu voltage sag fault with $\omega_{\text{pll}}=16~\text{Hz}\times2\pi$, $i_d=i_{\text{d,sep}}$.}
\label{exp_pll} 
\end{figure}

\begin{figure}[htbp]
\centering
\subfloat[Phase portraits in DVC state phase plane under $\omega_{\text{dvc}}$ maintained at $16~\text{Hz}\times2\pi$ (blue line) and $\omega_{\text{pll}}$ decreased to $0.4~\text{Hz}\times2\pi$ at 0.35~s (green line).]{\includegraphics[width=0.32\textwidth]{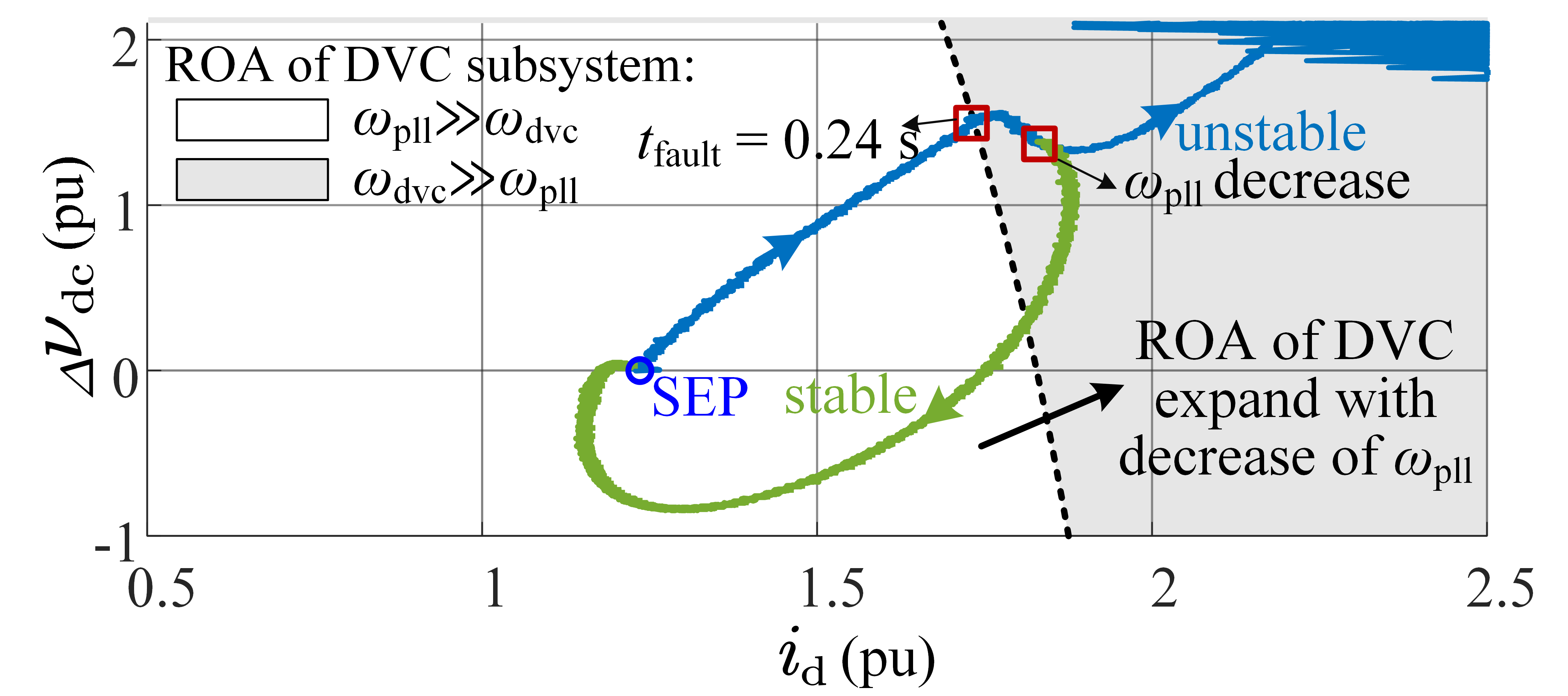}}\\ 
\subfloat[Time-domain results when $\omega_{\text{pll}}$ maintained at $16~\text{Hz}\times2\pi$.]{\includegraphics[width=0.46\textwidth]{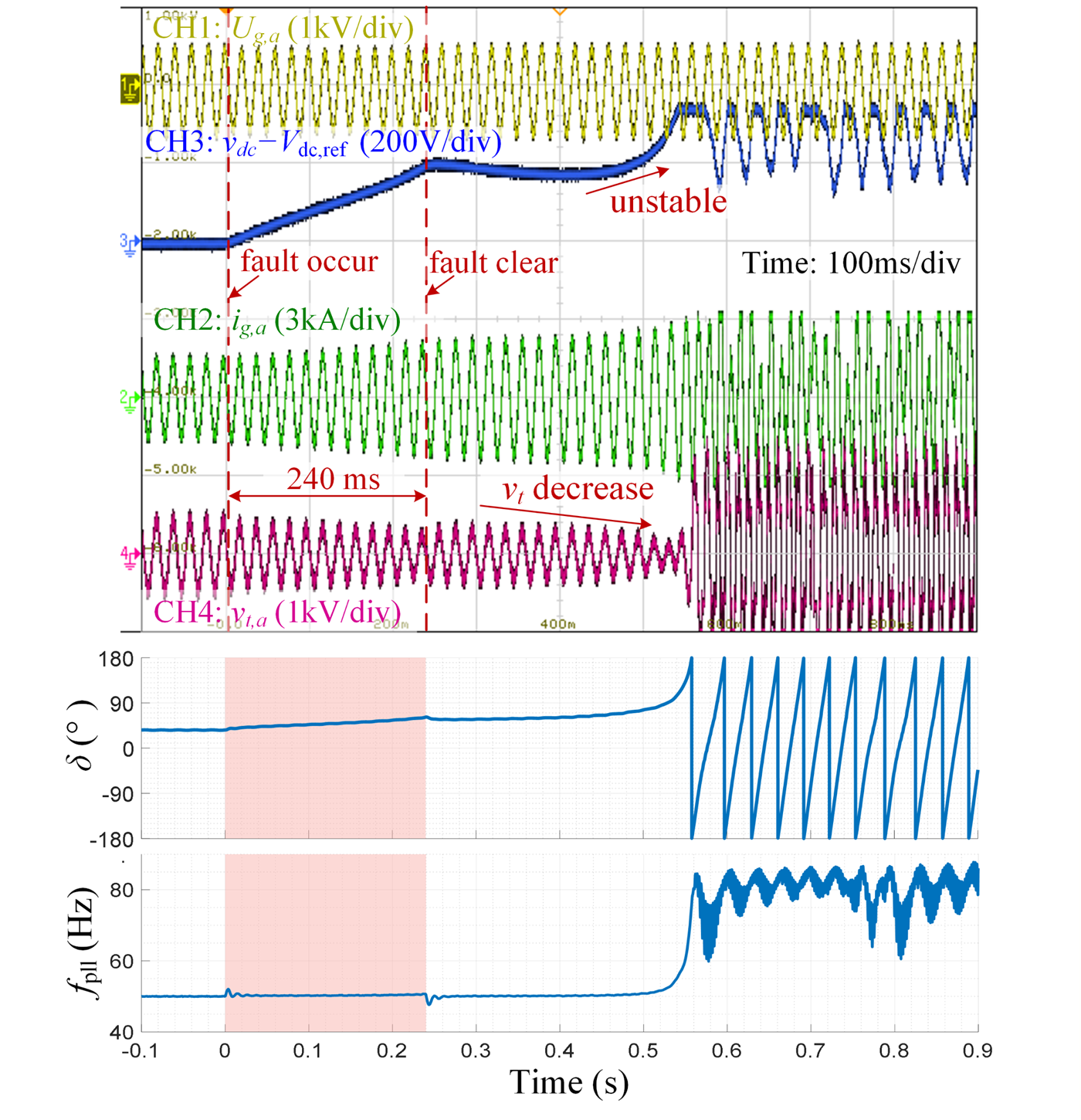}}\\
\subfloat[Time-domain results when $\omega_{\text{pll}}$ decreased to $0.4~\text{Hz}\times2\pi$ at 0.35~s.]{\includegraphics[width=0.46\textwidth]{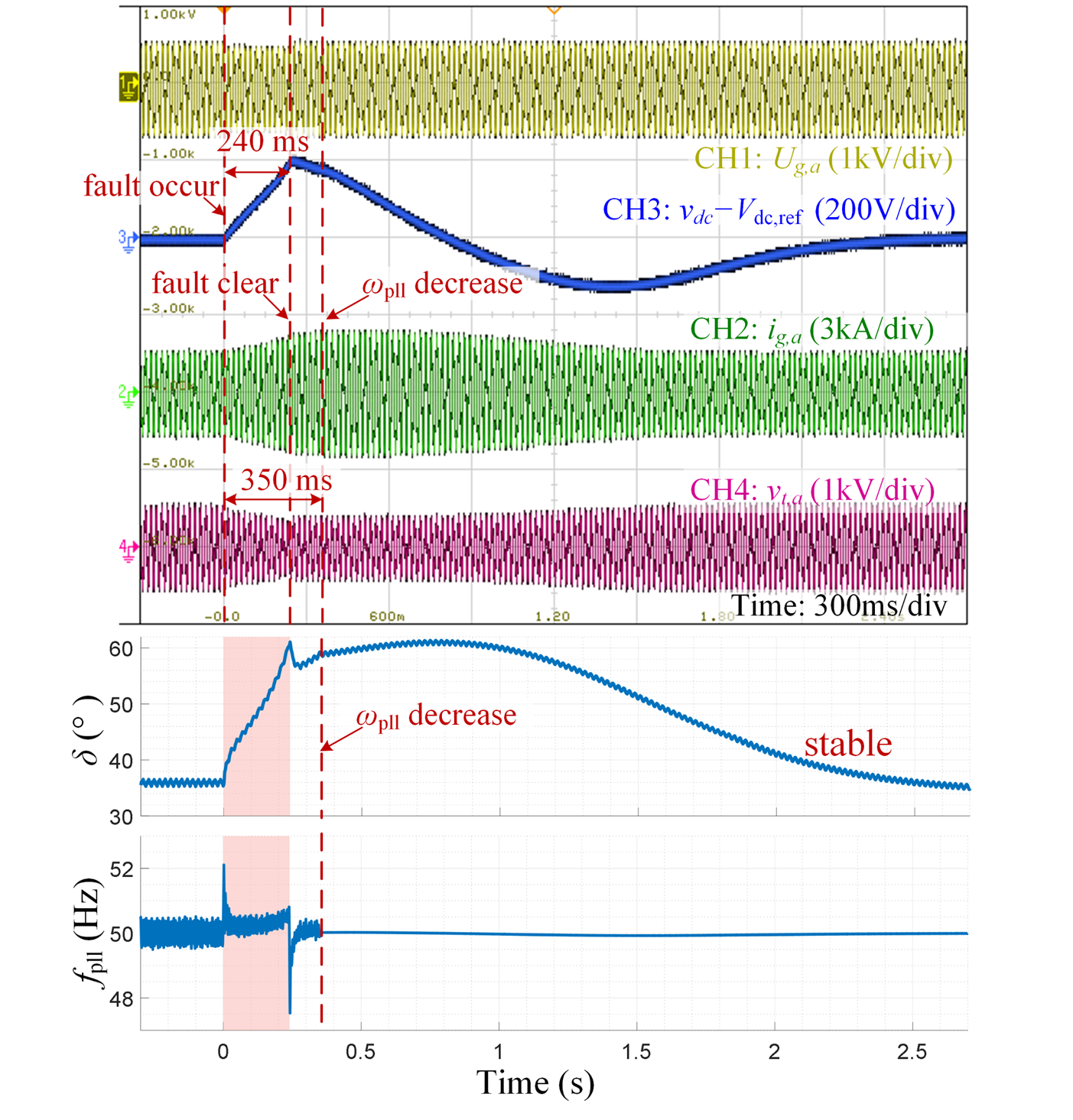}}\\
\caption{GFL inverter under 0.9 pu voltage sag fault with slow TVC $\omega_{\text{tvc}}=0.3~\text{Hz}\times2\pi$, $\omega_{\text{pll}}=16~\text{Hz}\times2\pi$, $\omega_{\text{dvc}}=2~\text{Hz}\times2\pi$, and fault clearing time $t_{\text{fault}}$=0.24~s.}
\label{exp_1_2}
\end{figure}

In \figref{exp_pll}, the system is PLL standalone, that is, there are no DVC or TVC control loops.  Under this configuration, the system remains stable even when the grid voltage drops to 0.6 pu without fault clearance. However, when both PLL and DVC loops are included, a 0.9 pu voltage sag under a weak grid [short circuit ratio (SCR) of 2.1 herein] results in the absence of an equilibrium point (EP), and the system diverges during the fault. This is because the existence condition for an EP becomes $U_g > \sqrt{2P_{\text{in}}X_g} = 0.97$ when both DVC and PLL exist, which is more restrictive than the standalone PLL case where $U_g > i_dX_g = 0.47$. For instance, in \figref{exp_1_2}~(b), if the DVC loop is considered and the fault clearing time exceeds $t_{\text{fault}}=0.24$~s, the system will lose stability. This indicates that it is usually not easy for the PLL alone to cause system instability, but the interaction between DVC and PLL is the main factor for the instability. The previous works \cite{8632731, fu2020large, hu2019large, he2019transient, wu2019design, ma2021generalized, zhao2020nonlinear} do not take this situation into account. 

In \figref{exp_1_2}, the DVC is added, and the 0.9 pu voltage sag fault occurs at 0~s and is cleared at 0.24~s. Even though the TVC is added, the system will finally destabilize if the fault is not cleared. This is because the bandwidth of TVC is too low for $i_q$ to be adjusted effectively before the system becomes unstable. Precisely, in \figref{exp_1_2}~(b) configuration, the bandwidth of DVC is 2~Hz much smaller than that of PLL, and the CCT is nearly 0.24~s. After the fault is cleared, the DVC loop becomes unstable first, eventually causing the PLL and the whole system to lose stability.

The mechanism of the instability of the DVC loop is the adverse interaction loop between $i_d$ and $\delta$. The DVC controller tends to inject more $i_d$ in an attempt to regulate the DC voltage. The increase of $i_d$ will cause $\delta$ increase, which will, however, decrease the terminal voltage $v_t$, even after the fault is cleared. The reduction of $v_t$ leads to a decrease in output power and a further increase in $i_d$. In \figref{exp_1_2}~(b), this adverse interaction loop leads to $v_t$ continuing to decline and $i_d$ increasing finally to its limit 3.1~kA. Once $i_d > U_g/X_g$, the PLL subsystem then also becomes unstable. The DC-link voltage finally oscillates around 2~kV due to the DC chopper circuit, and cannot be restored.

\begin{figure}[t!]
\centering
\includegraphics[width=0.46\textwidth]{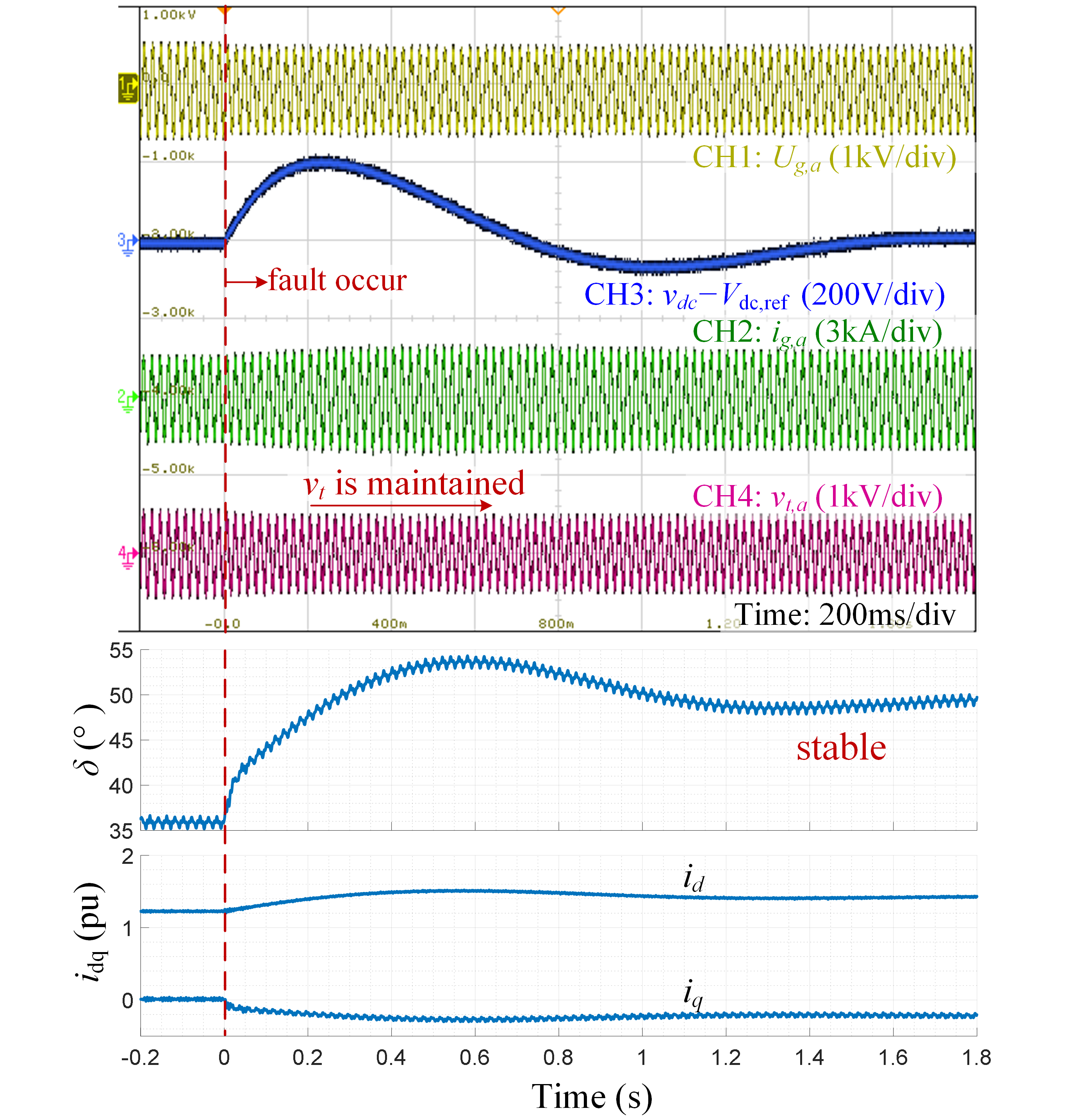}
\caption{GFL inverter under 0.9 pu voltage sag fault without fault clearance, with fast TVC $\omega_{\text{tvc}}=20~\text{Hz}\times2\pi$,$\omega_{\text{pll}}=16~\text{Hz}\times2\pi$,$\omega_{\text{dvc}}=2~\text{Hz}\times2\pi$, and without fault clear.}
\label{exp_1_3}
\end{figure}

\begin{figure}[htbp]
\centering
\subfloat[Phase portraits in DVC state phase plane under $\omega_{\text{dvc}}$ maintained at $15~\text{Hz}\times2\pi$ (blue line) and $\omega_{\text{dvc}}$ decreased to $0.4~\text{Hz}\times2\pi$ at 0.1~s when fault is cleared (green line). The red square represents the fault clearing point.]{\includegraphics[width=0.36\textwidth]{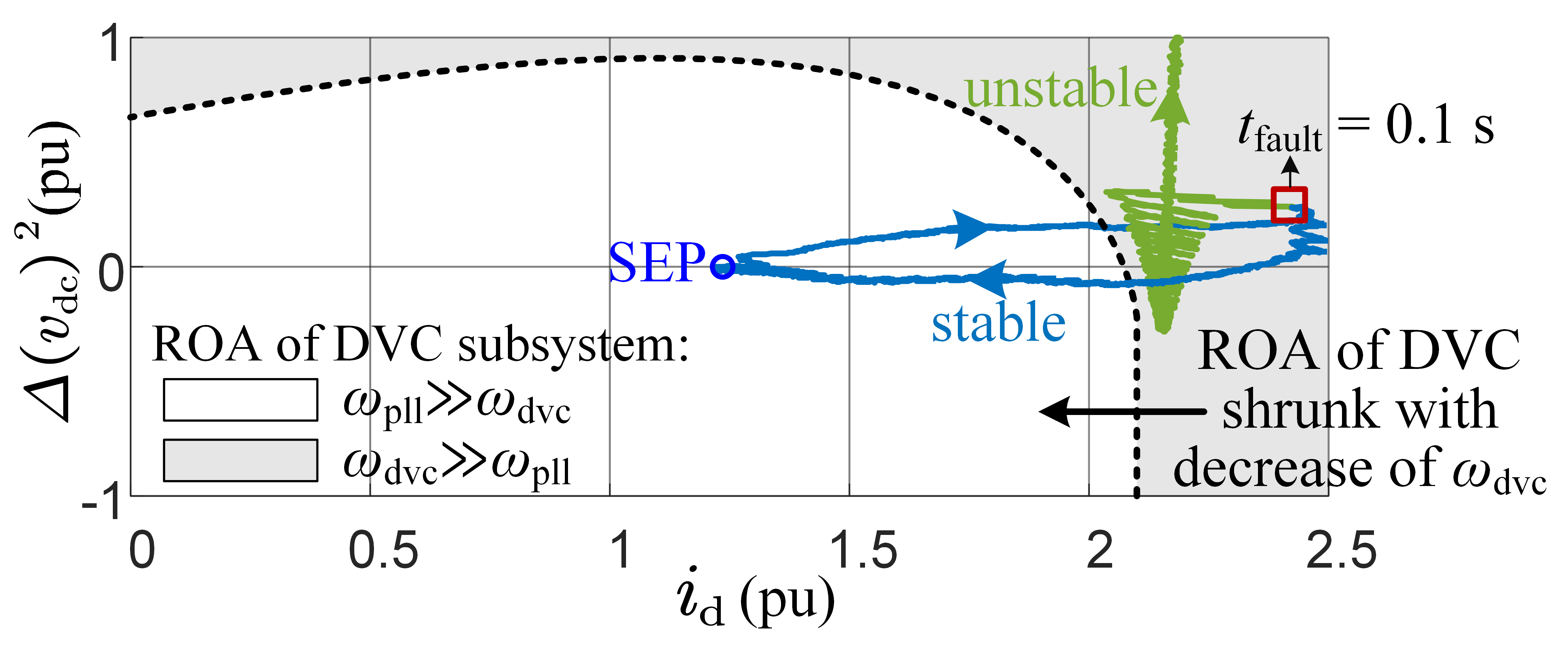}}\\ 
\subfloat[Time-domain results when $\omega_{\text{dvc}}$ maintained at $15~\text{Hz}\times2\pi$.]{\includegraphics[width=0.465\textwidth]{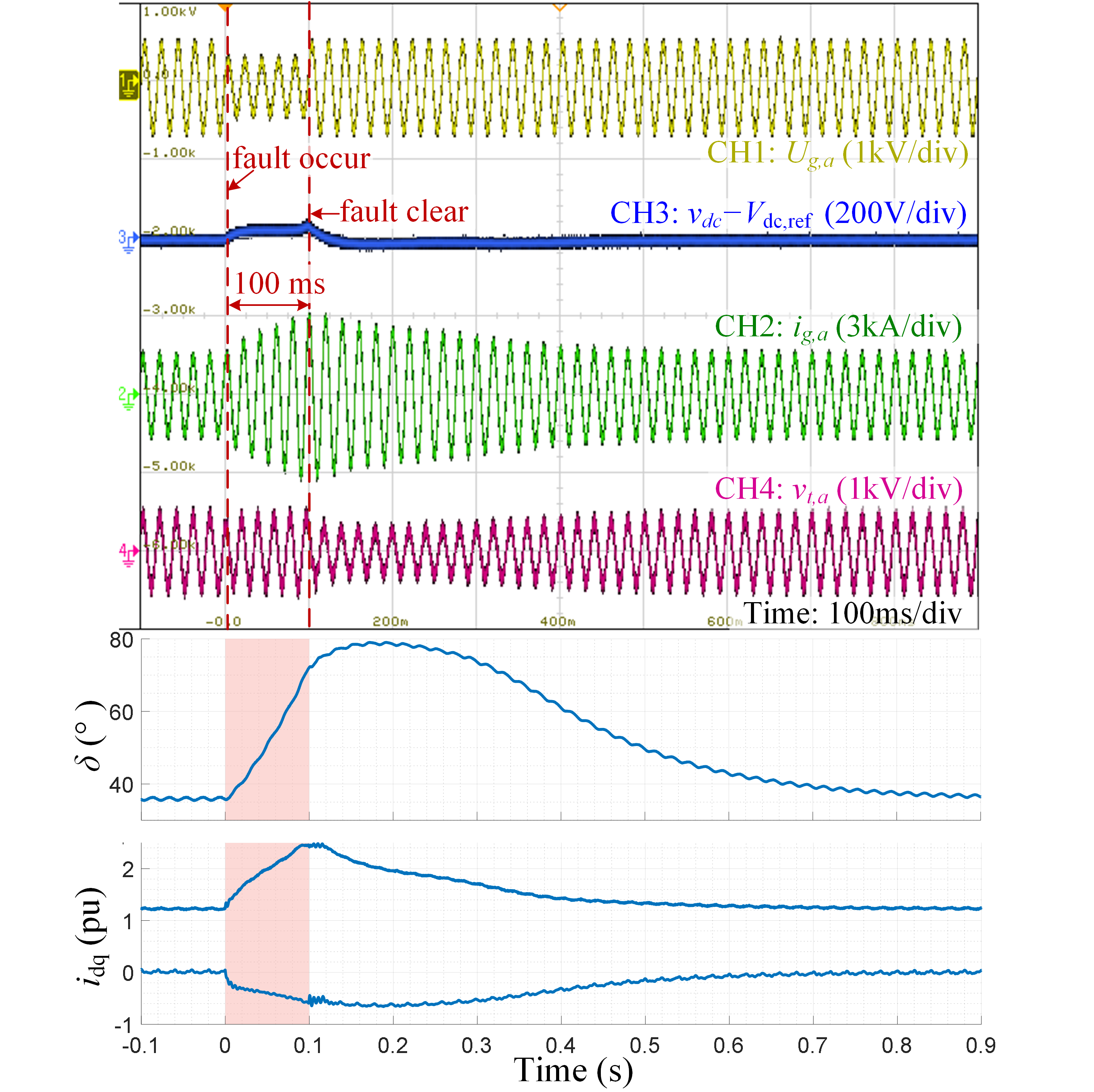}}\\
\subfloat[Time-domain results when $\omega_{\text{dvc}}$ decreased to $0.4~\text{Hz}\times2\pi$ at 0.1~s when fault is cleared.]{\includegraphics[width=0.474\textwidth]{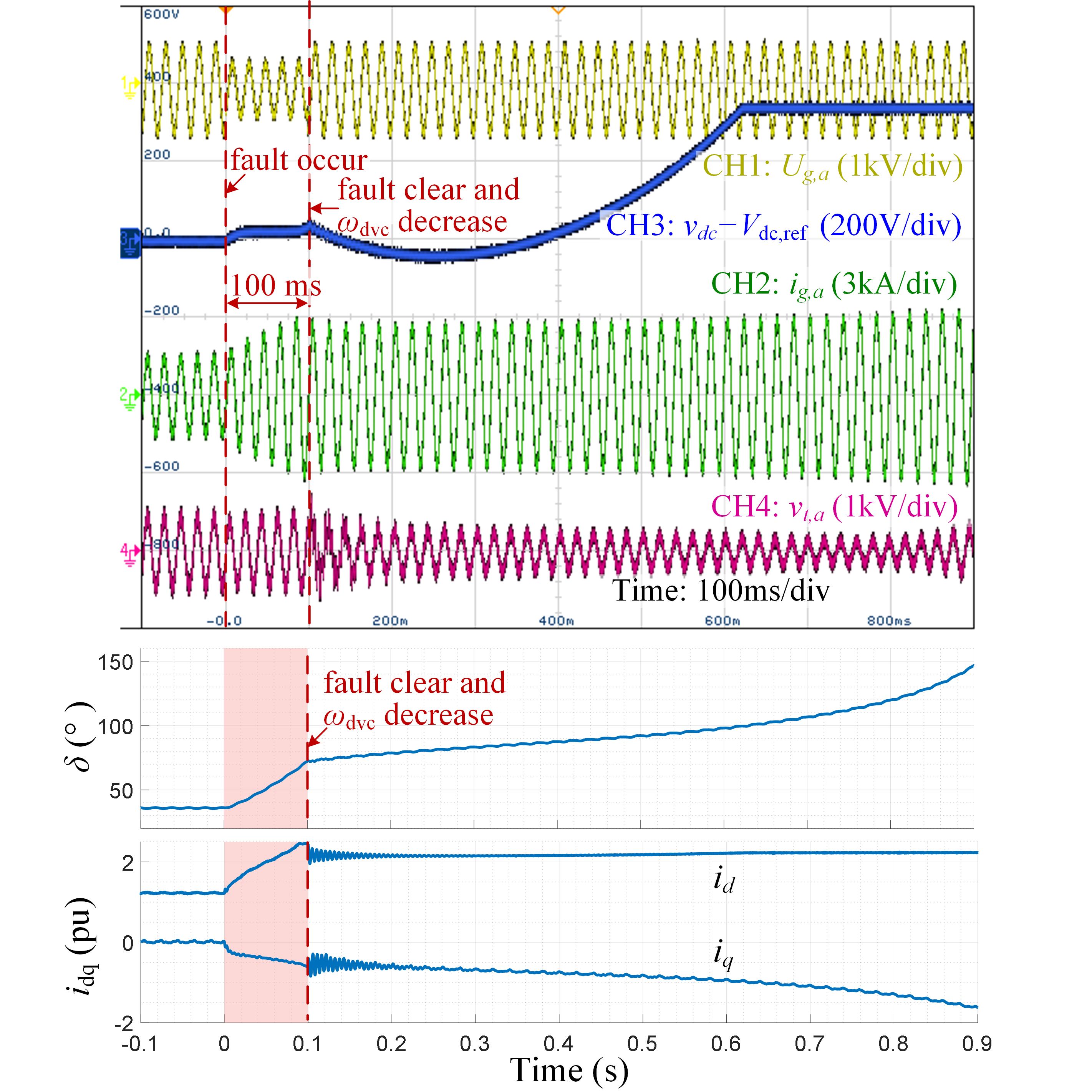}}\\
\caption{GFL inverter under 0.6 pu voltage sag fault with fast TVC $\omega_{\text{tvc}}=20~\text{Hz}\times2\pi$, $\omega_{\text{pll}}=2~\text{Hz}\times2\pi$, $\omega_{\text{dvc}}=15~\text{Hz}\times2\pi$, and fault clearing time $t_{\text{fault}}$=0.1~s.}
\label{exp_1_4}
\end{figure}

In \figref{exp_1_2}~(c), $\omega_{\text{pll}}$ is reduced to 0.4~Hz at 0.35~s after the fault is cleared. This renders DVC the fast subsystem and extends its stability region across the full phase plane, as seen in \figref{exp_1_2}~(a). Since $\delta$ remains within the ROA of the PLL subsystem at 0.35~s, and the DVC states are also within their new ROA, the system converges to its SEP. The reduction in $\omega_{\text{pll}}$ limits the rapid growth of $\delta$, thereby interrupting the adverse interaction that would otherwise reduce $v_t$.
 
Compared to \figref{exp_1_2}~(b), the change in \figref{exp_1_3} is the increase of TVC bandwidth to 20~Hz, which is significantly higher than DVC. This allows $i_q$ to respond rapidly before the system becomes unstable, maintaining overall stability even without fault clearance. The injected $i_q$ helps sustain the terminal voltage $v_t$. As discussed in Section~III, when $k_v \to \infty$, $v_t$ becomes tightly regulated to $V_{\text{ref}}$, and the increase of $\delta$ due to $i_d$ no longer reduces power output. This breaks the adverse interaction loop. Therefore, voltage instability is the fundamental cause of instability in GFL inverters with DVC. A higher TVC bandwidth improves $v_t$ regulation and contributes to transient stability. This benefit is also consistent with the small-signal analysis in \cite{huang2017effect}.

In \figref{exp_1_4}, the bandwidths of PLL, DVC, and TVC are set to 2~Hz, 15~Hz, and 20~Hz respectively, which is the same as the configuration in \figref{Sim_tvc}~(b). The UEP is located at $\delta_{\text{uep}}=80.3^\circ$ and the CCT under 0.6~pu voltage sag fault is 0.1~s. In \figref{exp_1_4}~(b), the DVC loop is significantly faster than the PLL. When the fault is cleared at $t_{\text{fault}} = 0.1$~s, the system remain critically stable with $\delta$ staying below $\delta_{\text{uep}}$. The trajectory projected onto the DVC state phase plane is plotted in \figref{exp_1_4}~(a). Since the DVC acts as the fast subsystem, its states can stabilize over the entire phase plane as long as the slow PLL remains stable. By contrast, in \figref{exp_1_4}~(c), the DVC bandwidth is reduced to 0.4~Hz, making it the slowest control loop. At the fault clearing point, the projected trajectory in \figref{exp_1_4}~(a) shows $i_d$ exceeding 2pu, which lies outside the ROA of the DVC subsystem in its slow configuration. Consequently, the system under slow DVC configuration loses stability. The underlying mechanism is as follows: when $\omega_{\text{dvc}}$ is high, $i_d$ can respond rapidly to regulate the DC voltage, with $\delta$ remaining relatively constant. However, when $\omega_{\text{dvc}}$ is low, the PLL becomes comparatively faster. In this case, the increase in $i_d$ leads to a rapid rise in $\delta$, which in turn reduces $v_t$ and the output power. As illustrated in \figref{exp_1_4}~(c), this destabilizes the DVC control loop, resulting in a DC voltage rise toward its upper limit and eventual system instability.

\begin{figure}[t!]
\centering
\includegraphics[width=0.485\textwidth]{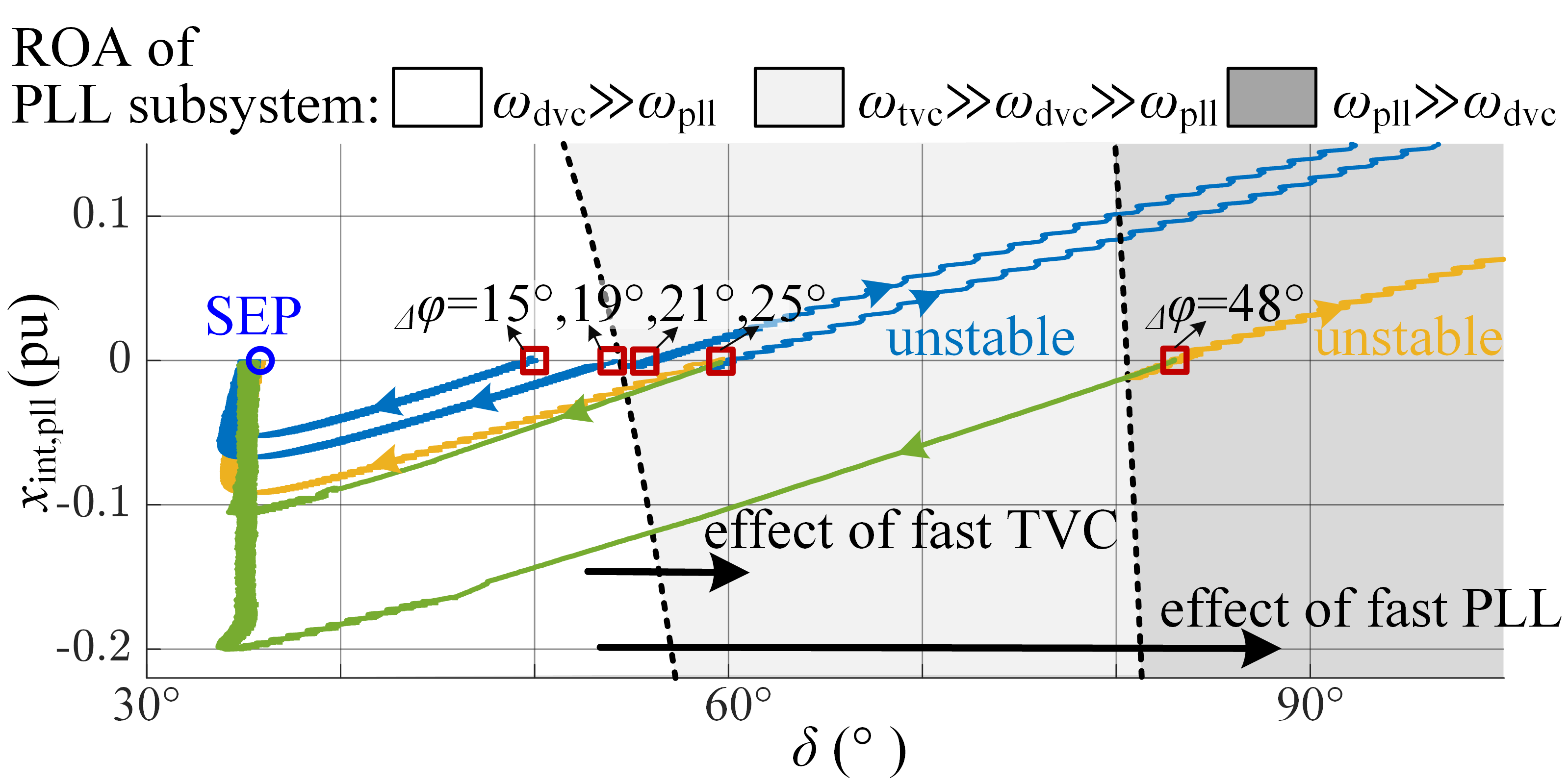}
\caption{Phase portraits of phase jump fault under ${}_{\Delta}\varphi=$15, 19, 21, 25, 48 degrees. Blue trajectories: $\omega_{\text{pll}}=1~\text{Hz}\times2\pi$, $\omega_{\text{dvc}} = 8~\text{Hz}\times2\pi$, no TVC; green trajectories: $\omega_{\text{pll}}$ increased to $20~\text{Hz}\times2\pi$; orange trajectories: fast TVC is added with $\omega_{\text{tvc}} = 20~\text{Hz}\times2\pi$, $\omega_{\text{pll}}$ remained at $1~\text{Hz}\times2\pi$.}
\label{exp_2_1}
\end{figure}

\begin{figure}[t!]
\centering
\subfloat[Bnadwith of PLL is maintained at 1~Hz.]{\includegraphics[width=0.45\textwidth]{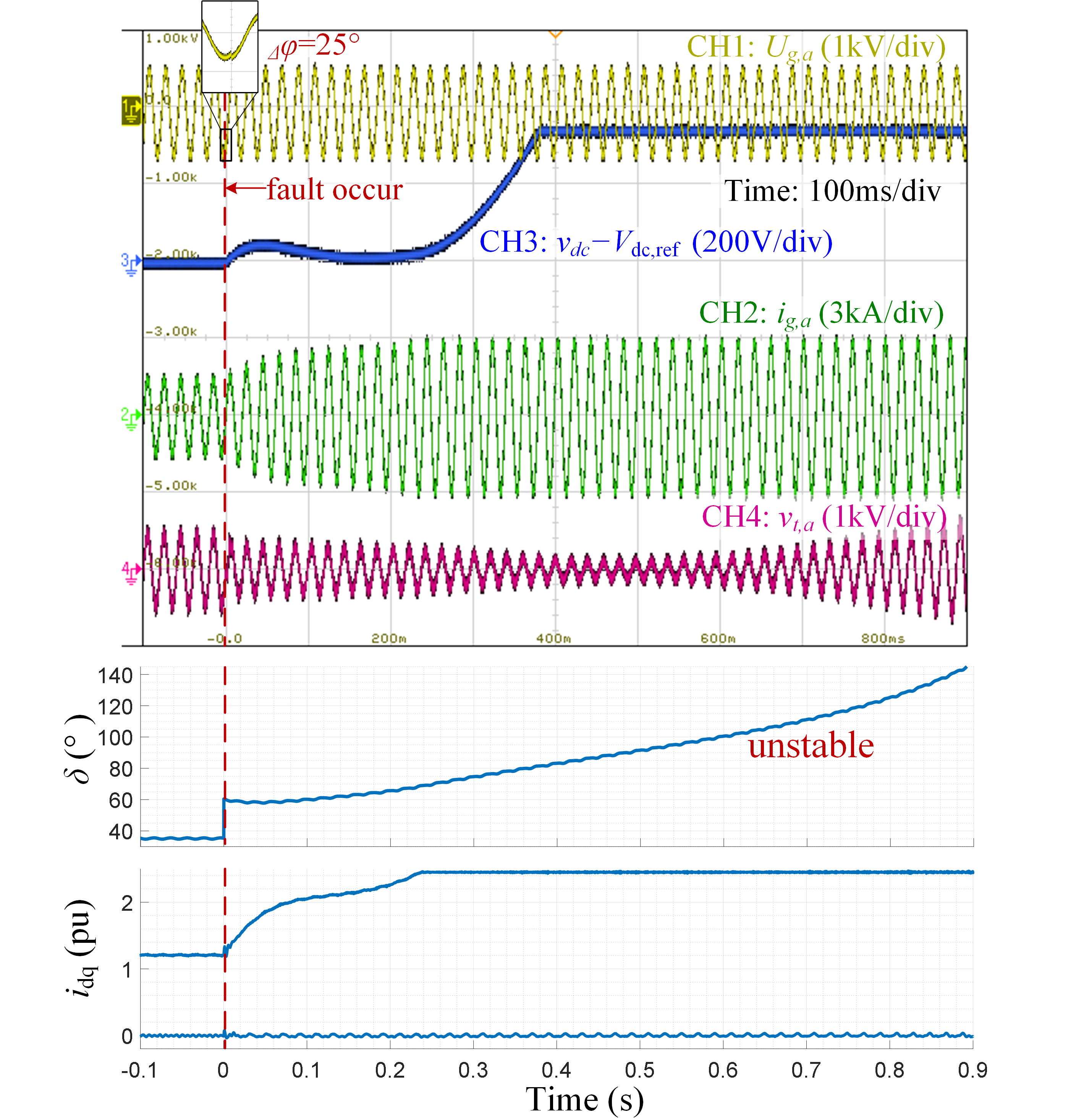}}\\ 
\subfloat[Bnadwith of PLL is increased to 20~Hz.]{\includegraphics[width=0.45\textwidth]{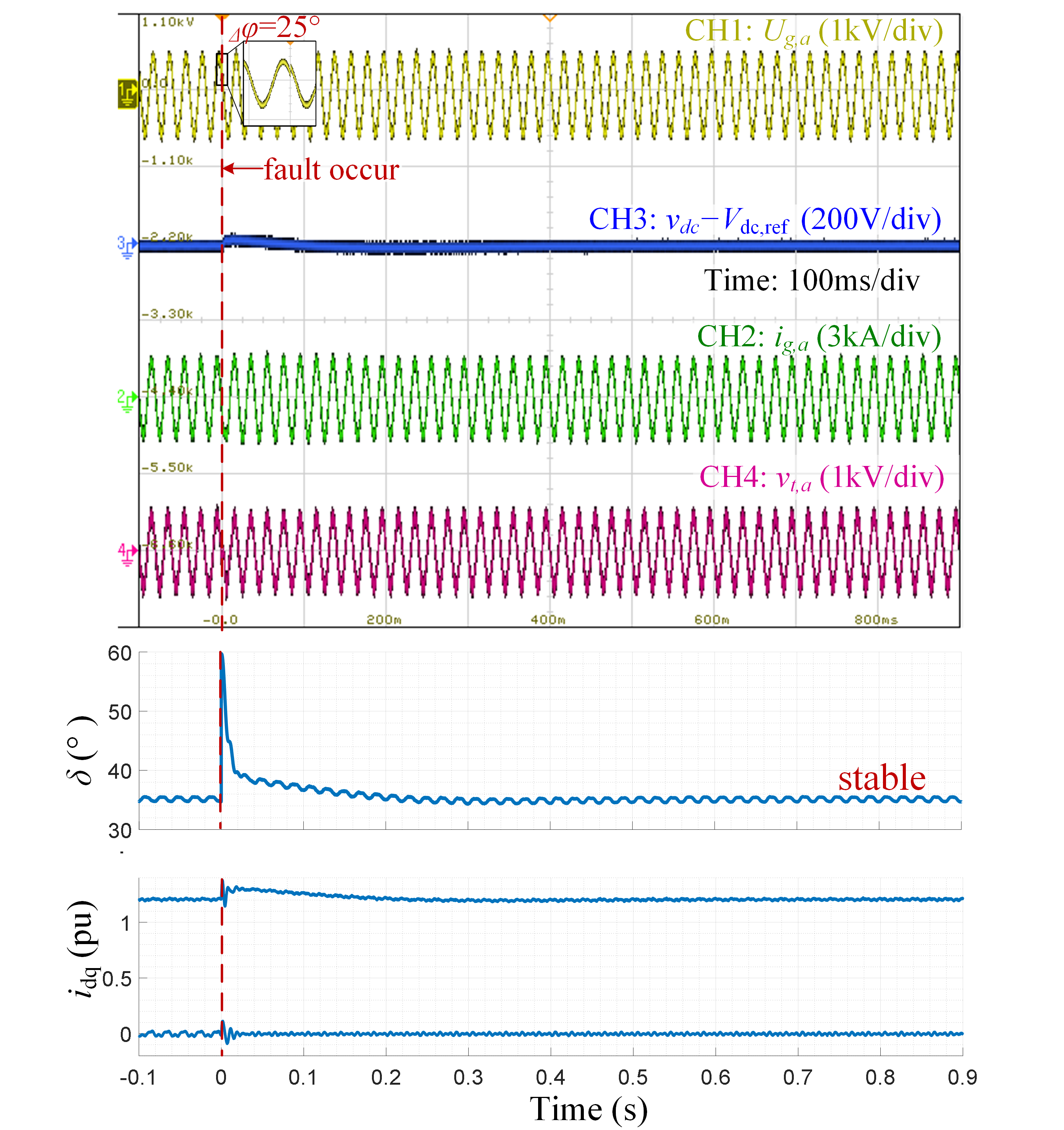}}\\
\caption{Time-domain results for GFL inverter under 25~degree phase jump fault with $\omega_{\text{dvc}}=8\times2\pi$.}
\label{exp_2_2}
\end{figure}

\subsection{Phase jump fault}
This section investigates the system response under phase jump disturbances. The bandwidths of the PLL and DVC are initially set to 1Hz and 8Hz, respectively, placing the PLL as the slowest control loop. According to the previous analysis, this configuration yields a UEP at $\delta_{\text{uep}} = 54^\circ$, corresponding to the smallest stability region in \figref{exp_2_1}. When the phase jump value ${}_{\Delta}\varphi$ is $15^\circ$, $19^\circ$, $21^\circ$, $25^\circ$, and $48^\circ$, the system trajectories are plotted in \figref{exp_2_1} as blue lines. These results align with the theoretically derived ROA of the reduced-order system under the $\omega_{\text{dvc}} \gg \omega_{\text{pll}}$ configuration.

By increasing the PLL bandwidth to 20Hz, the system can remain stable even under a $48^\circ$ phase jump as shown by green trajectories in \figref{exp_2_1}. \figref{exp_2_2} compares time-domain simulations under a $25^\circ$ phase jump with $\omega_{\text{pll}}$ at 1Hz and 20Hz, respectively. In \figref{exp_2_2}~(a), due to the relatively high bandwidth of DVC, $i_d$ rises rapidly following the fault in an attempt to regulate output power and DC voltage, which leads to a subsequent increase in $\delta$. This, in turn, further boosts $i_d$, eventually pushing it to its limit of 2.5~pu, thereby causing the system to destabilize. In contrast, when the PLL is faster, as shown in \figref{exp_2_2}~(b), $\delta$ is quickly adjusted by the PLL first, allowing the DVC then to follow the correct regulation. As a result, the system remains stable. Alternatively, even without modifying the PLL bandwidth, adding a fast TVC loop can expand the ROA of the DVC subsystem. However, this improvement is limited, with the UEP boundary extended only to $\delta_{\text{uep}}=80^\circ$. In \figref{exp_2_1}, the system becomes marginally unstable for a $48^\circ$ phase jump as evidenced by the orange trajectory. In comparison, simply increasing the PLL bandwidth, without TVC, significantly enlarges the stability region, theoretically extending beyond $\delta_{\text{uep}} > 90^\circ$ and approaching that of a standalone PLL system.

It is also important to note that the ability to suppress phase jump disturbances depends primarily on the relative speeds of the PLL and DVC loops. Similar stabilizing effects can be achieved by reducing the DVC bandwidth either. Overall, a relatively fast PLL loop is more effective in enhancing transient stability under phase jump conditions.

\subsection{Input power variation}
\begin{figure}[t!]
\centering
\includegraphics[width=0.485\textwidth]{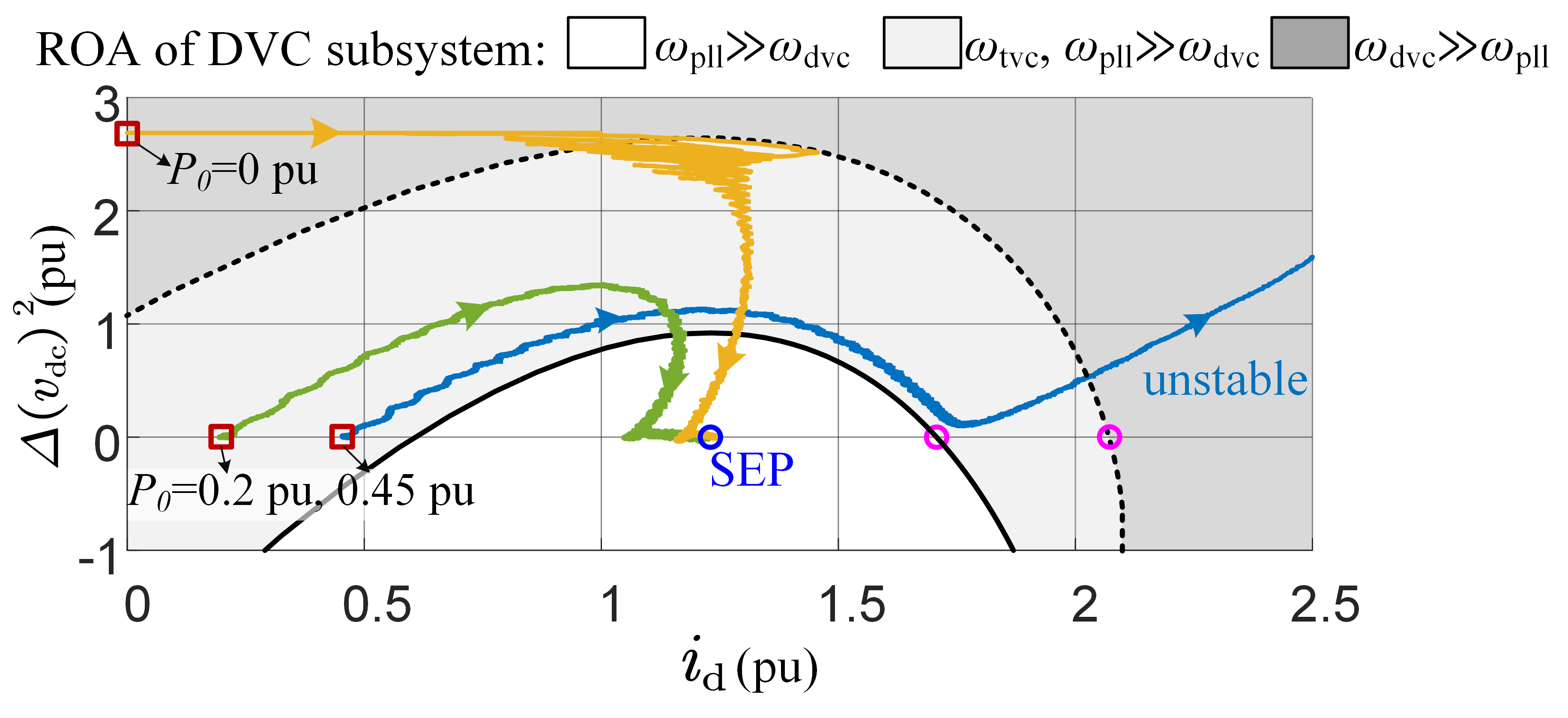}
\caption{Phase portrait under input power step changes from 0, 0.2~pu, and 0.45~pu to 1~pu. 
Blue trajectories: $\omega_{\text{pll}} = 18~\text{Hz} \times 2\pi$, $\omega_{\text{dvc}} = 3~\text{Hz} \times 2\pi$, no TVC. 
Green trajectories: $\omega_{\text{pll}}$ decreased to $0.5~\text{Hz} \times 2\pi$, $\omega_{\text{dvc}}$ unchanged, no TVC. 
Orange trajectories: $\omega_{\text{pll}}$ decreased to $0.5~\text{Hz} \times 2\pi$, $\omega_{\text{dvc}}$ unchanged, with fast TVC added at $\omega_{\text{tvc}} = 20~\text{Hz} \times 2\pi$.}
\label{exp_3_1}
\end{figure}

\begin{figure}[t!]
\centering
\subfloat[Input power step from $P_{\text{in}} = 0.45$~pu to 1~pu with $\omega_{\text{pll}} = 18~\text{Hz} \times 2\pi$, $\omega_{\text{dvc}} = 3~\text{Hz} \times 2\pi$.]{
    \includegraphics[width=0.46\textwidth]{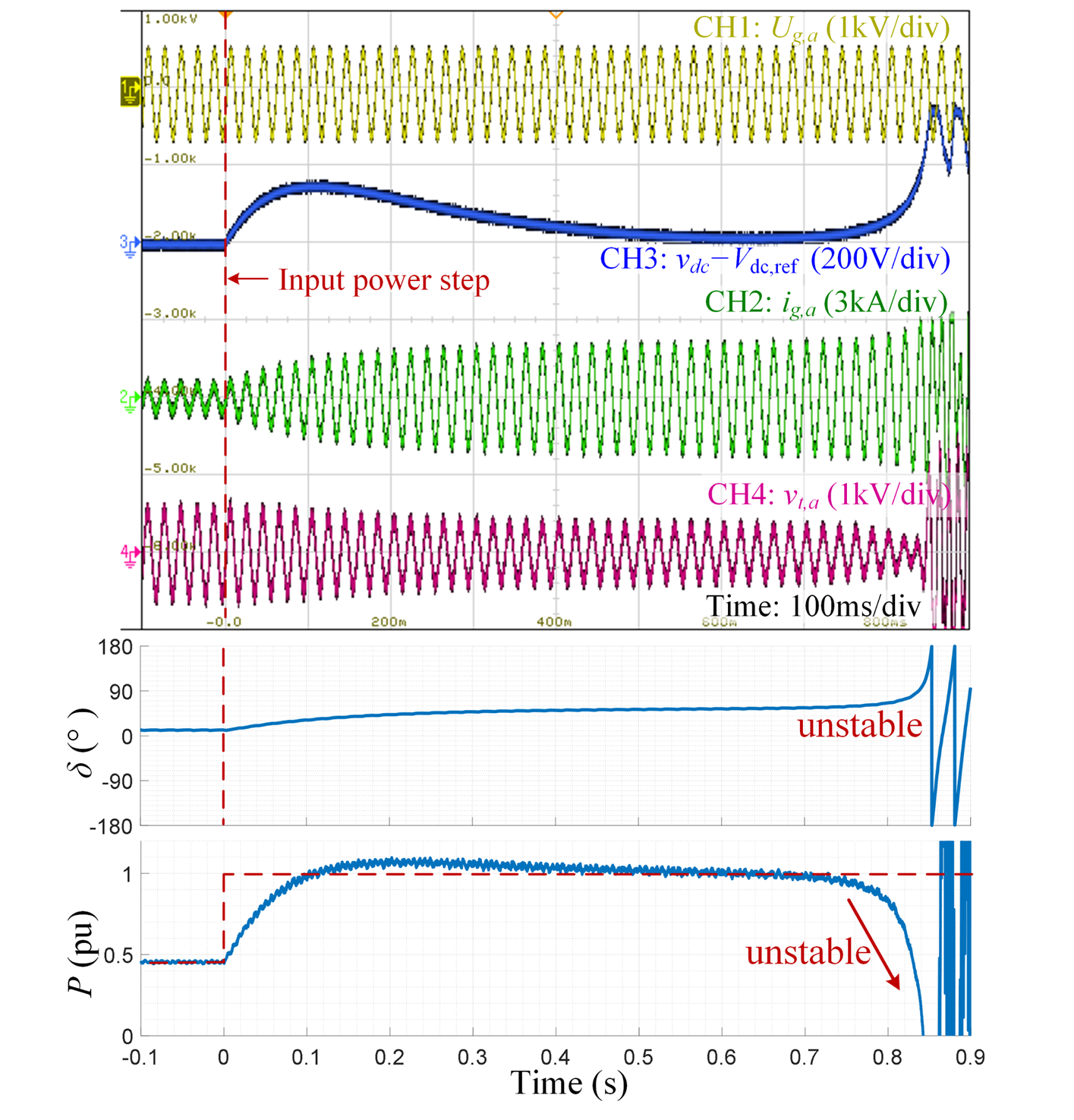}
}\\
\subfloat[Input power step from $P_{\text{in}}  = 0.2$~pu to 1~pu with $\omega_{\text{pll}}$ reduced to $0.4~\text{Hz} \times 2\pi$, while $\omega_{\text{dvc}}$ remains at $3~\text{Hz} \times 2\pi$.]{
\includegraphics[width=0.46\textwidth]{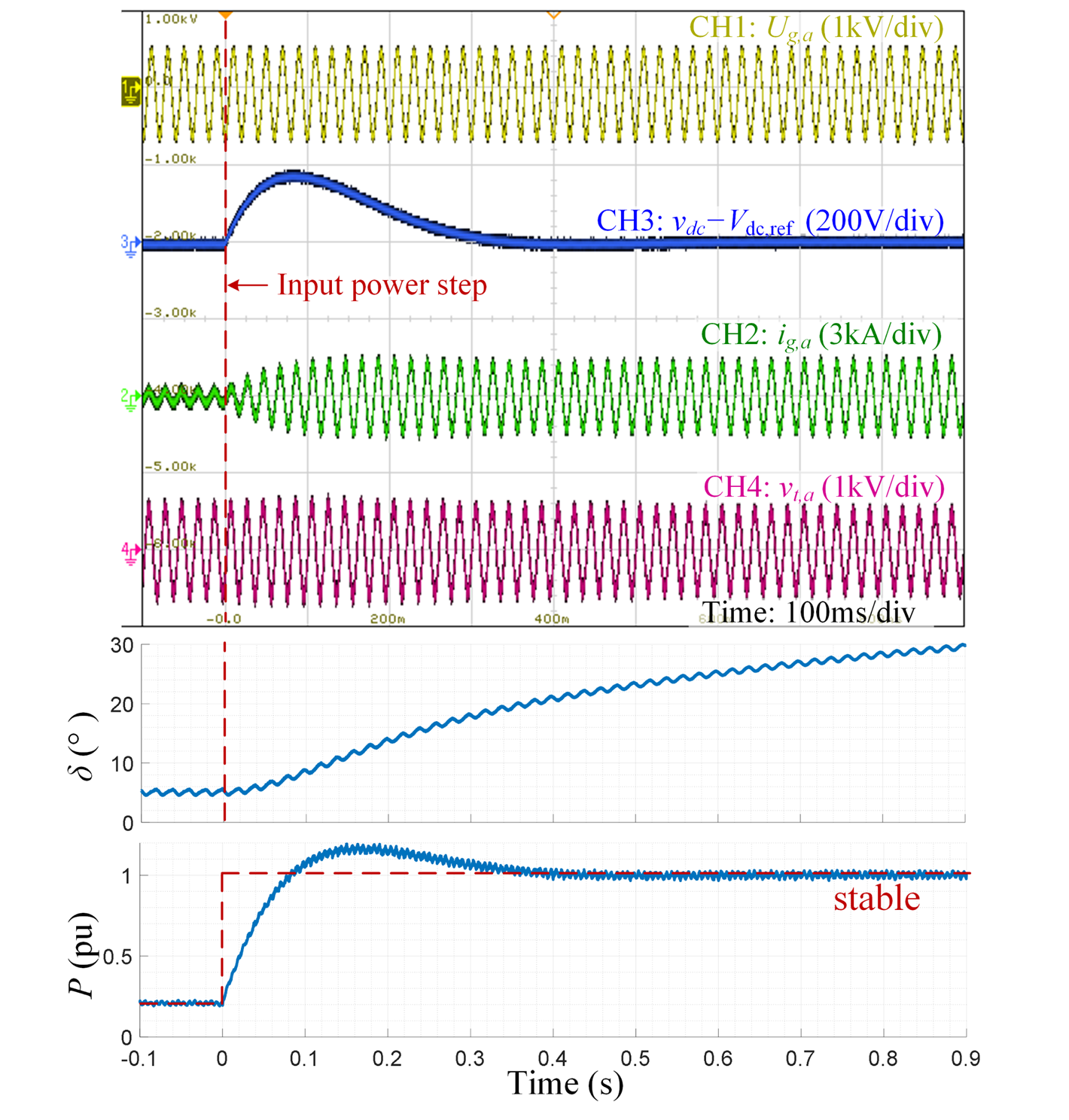}
}\\
\caption{Time-domain results of the GFL inverter under input power step changes.}
\label{exp_3_2}
\end{figure}

\begin{figure}[t!]
\centering
\includegraphics[width=0.46\textwidth]{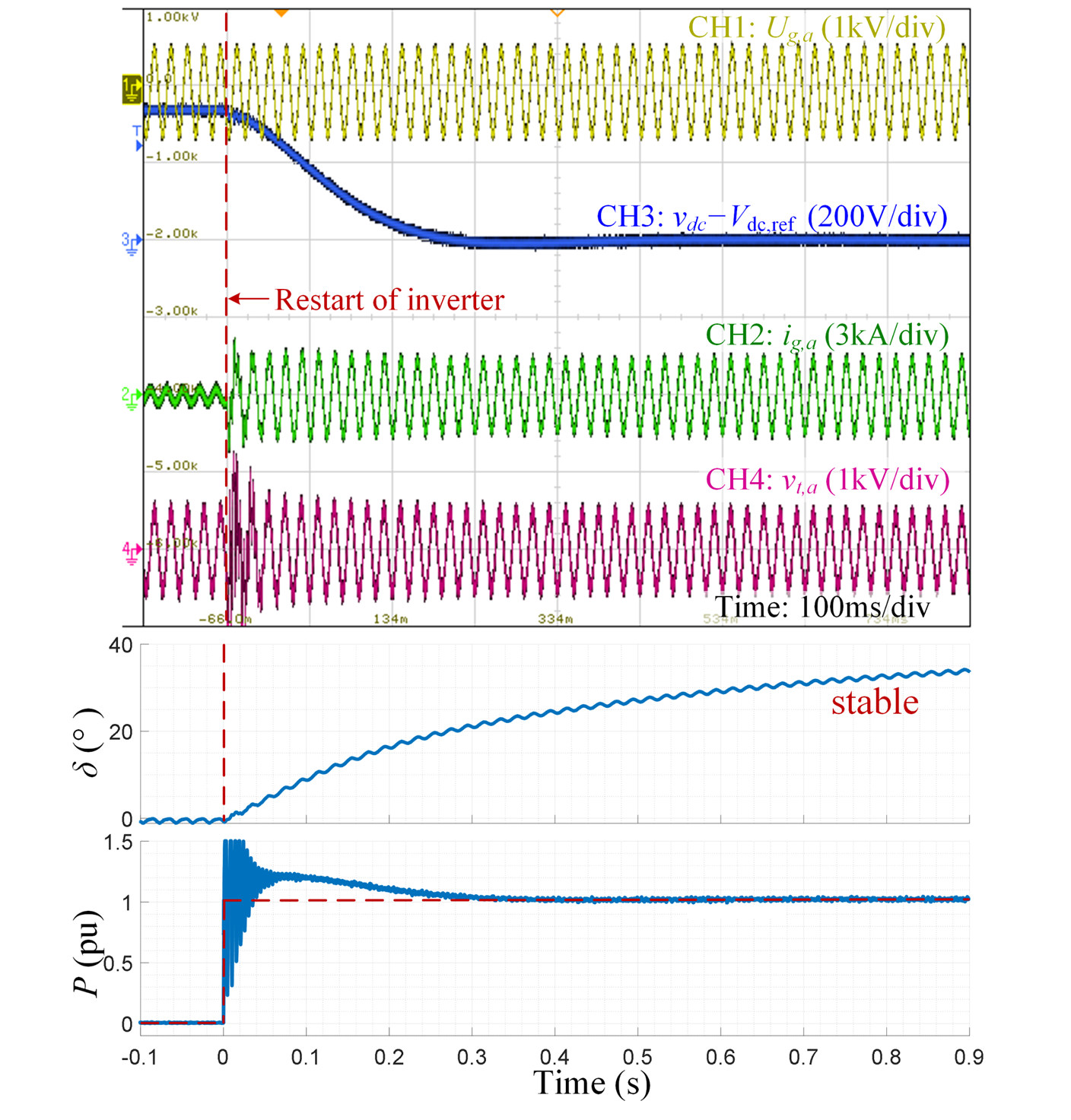}
\caption{Time-domain response of the GFL inverter during restart. The output power steps from 0 to 1~pu. Control parameters: $\omega_{\text{pll}} = 0.5~\text{Hz} \times 2\pi$, $\omega_{\text{dvc}}$ unchanged, and fast TVC added with $\omega_{\text{tvc}} = 20~\text{Hz} \times 2\pi$.}
\label{exp_3_3}
\end{figure}

This section analyzes synchronization loss caused by large power steps. Initially, the PLL and DVC bandwidths are set to 18~Hz and 3~Hz, respectively, i.e., $\omega_{\text{pll}} \gg \omega_{\text{dvc}}$. According to the previous analysis, this configuration corresponds to the smallest ROA for the reduced-order system in the DVC state phase plan in \figref{exp_3_1}. When input power steps from 0.45~pu to 1~pu, the original SEP lies outside the new SEP's ROA, leading to transient instability illustrated by the blue trajectory in \figref{exp_3_1}. This behavior is further illustrated in the time-domain response in \figref{exp_3_2}~(a), where the GFL inverter attempts to increase $i_d$ to maintain DC voltage. However, due to the fast PLL, the increase in $\delta$ leads to power delivery failure. Around 0.7~s, the inverter can no longer sustain 1~pu power output, resulting in rapid terminal voltage saturation and subsequent system instability. This phenomenon reflects large-signal instability, differing from small-signal cases reported in \cite{wu2020impact}. From the voltage response, the volatege is "first swing" stable but eventually unstable, and this instability mode is not captured in \cite{priyamvada2020online, 10567400}.

When the PLL bandwidth is reduced to 0.4~Hz while keeping the bandwidth of DVC unchanged, the system achieves the largest ROA in the DVC state phase plane. In this configuration, the system can remain stable when the input power steps from 0.2~pu to 1~pu, as illustrated by green trajectories in \figref{exp_3_1} and the corresponding time-domain results in \figref{exp_3_2}~(b). The relatively slower PLL alows the DVC to regulate $i_d$ before $\delta$ significantly increases, and then the PLL slowly tracks the new SEP. Additionally, fast TVC can also increase ROA in the DVC phase plane, as seen in \figref{exp_3_1}, but the improvement is limited compared to directly reducing PLL bandwidth or increasing DVC bandwidth. Even with TVC, stability cannot be guaranteed for more severe disturbances such as inverter restarts.

After major grid faults, the grid-side inverter may shut down and later restart. During restart, the DC voltage is initially near its upper limit (e.g., 3~pu) due to the chopper circuit, and the output power jumps from 0 to 1~pu. If the PLL bandwidth remains high, instability occurs even with fast TVC during restart, because the initial point lies outside the ROA of SEP. In \figref{exp_3_1}, with $\omega_{\text{pll}}$ reduced to 0.4~Hz and TVC added, the inverter trajectory (orange line) starts from $i_d = 0$, $\varDelta \left( v_{\mathrm{dc}} \right) ^2 = v_{\text{dc}}^2-V_{\text{dc,ref}}^2=2.75$ and converges to SEP. The time-domain results are shown in \figref{exp_3_3}. The overall control sequence is that the PLL converges first, and then its speed slows down, after which the DVC is added to let the DC voltage move to the SEP.

\section{Conclusions}
This paper proposes a \textit{bandwidth separation method} to analyze the transient stability of the GFL inverter considering the interaction among three outer control loops: PLL, DVC, and TVC. By using this asymptotic analysis approach, the model order can be reduced, which provides physical insight into the nonlinear dynamics under large disturbances. The ROA for each subsystem can also be derived systematically under sufficient bandwidth separation. The analysis reveals that the interaction between PLL and DVC loops significantly degrades system stability and reduces the overall stability region. More specifically, a relatively large PLL bandwidth narrows the ROA in the DVC phase plane and can induce voltage instability through adverse interaction with $i_d$. Conversely, a large DVC bandwidth relative to the PLL bandwidth narrows the ROA of the PLL subsystem. A fast TVC loop, when configured with a higher bandwidth than the DVC, can effectively mitigate this adverse interaction between PLL and DVC, and expand the stability region in both PLL and DVC subsystems.

Under voltage sag conditions, the root cause of instability is identified as voltage instability and a subsequent output power capacity reduce, arising from the aforementioned adverse interaction between PLL and DVC. This mechanism highlights that GFL inverters are more vulnerable to losing stability than what conventional PLL-only analyses suggest, a factor often overlooked in prior studies. Incorporating a fast TVC loop is shown to weaken this destabilizing interaction and enhance the inverter’s output power capability under transient stability. Moreover, the study identifies optimal bandwidth configurations under different disturbance types. A fast PLL improves resilience to phase jumps, while a slower PLL or a faster DVC is more effective in handling power variations and restart scenarios. The addition of a fast TVC loop further improves stability by reinforcing terminal voltage regulation and alleviating the PLL–DVC adverse interaction effects.

All theoretical results are validated through HIL experiments and simulations, confirming the accuracy of the reduced models and the findings of the proposed analysis method. Even though sometimes the control loop bandwidths may not be ideally separated in all inverters, the proposed analysis method can still indicate trends in characteristics and guide the design choices of inverter control loop bandwidths. Moreover, this analysis framework and the derived reduced-order GFL model can be used on multi-inverter systems in future work.

\vspace{-0.1cm}
\appendices
\setcounter{table}{0}
\setcounter{figure}{0}
\renewcommand{\thetable}{A\arabic{table}}
\renewcommand\cellalign{tl}
\renewcommand{\theequation}{A\arabic{equation}}
\renewcommand{\thefigure}{A\arabic{figure}}
\section{Full-Order ODE Model of GFL Inverter} \label{A}
In this paper, the PLL is designed to be over-damped, while the DVC and the inner current loop are implemented as typical second-order systems.

The PI controllers are designed in the small-signal domain. The closed-loop transfer function of the PLL is given by
\[
G_{\mathrm{pll}}(s) = \frac{{}_{\varDelta} \delta}{{}_{\varDelta} V_q} = \frac{k_{\mathrm{p,pll}} + k_{\mathrm{i,pll}}/s}{s + V_d(k_{\mathrm{p,pll}} + k_{\mathrm{i,pll}}/s)}
\]
according to \cite{wen2015analysis}. Assuming $V_d \approx 1$ at steady state and neglecting the integral term, which corresponds to an overdamped PLL design, we have $\left. G_{\mathrm{pll}}(j\omega_{\mathrm{pll}}) \right|_{\omega_{\mathrm{pll}} = k_{\mathrm{p,pll}}} = -3$~dB. This overdamped setting is commonly adopted since the small integral term and large proportional term are beneficial to transient stability \cite{hu2019large,wu2019design,li2022whole}. In this paper, we define $\zeta_{\mathrm{pll}} = k_{\mathrm{i,pll}} / k_{\mathrm{p,pll}}$.

The closed-loop transfer function of the DVC is:
\[
G_{\mathrm{dvc}}(s) = \frac{k_{\mathrm{p,dvc}} + k_{\mathrm{i,dvc}}/s}{sC_{\mathrm{dc}}/\omega_s + k_{\mathrm{p,dvc}} + k_{\mathrm{i,dvc}}/s}
\]
and the gains are selected as $k_{\mathrm{p},\mathrm{dvc}}=\frac{C_{\mathrm{dc}}/\omega _s}{2}\cdot \omega _{\mathrm{dvc}}$ and $k_{i,\mathrm{dvc}}=\zeta _{\mathrm{dvc}}\frac{C_{\mathrm{dc}}/\omega _s}{2}\cdot {\omega _{\mathrm{dvc}}}^2$. When $\zeta_{\mathrm{dvc}} = 1/4$, the DVC becomes a critically damped second-order system~\cite{li2022revisiting}, with $G_{\mathrm{dvc}}(j\omega_{\mathrm{dvc}}) \approx -3$~dB. The same PI design approach is also applied to the inner current loop.

By substituting the above expressions for loop bandwidths, the nonlinear differential equations in \eqref{PLL},\eqref{DVC}, and \eqref{TVC} can be reformulated as \eqref{ODE}, and the detailed expressions in \eqref{ODE} are as follows:
\small
\begin{subequations}
\begin{equation}
\begin{aligned}
f_{\mathrm{pll},1}\left( \delta , i_d, \frac{x_{\mathrm{int},\mathrm{pll}}}{\omega _{\mathrm{pll}}} \right) = & \frac{X_gi_d-U_g\sin \delta +x_{\mathrm{int,pll}}/\omega _{\mathrm{pll}}}{1-k_{\mathrm{p},\mathrm{pll}}L_g/\mathrm{\omega}_{\mathrm{s}}\cdot i_d} \\
f_{\mathrm{pll},2}\left( \delta,i_d,x_{\mathrm{int},\mathrm{pll}} \right) = & \zeta _{\mathrm{pll}} \frac{X_gi_d-U_g\sin \delta +x_{\mathrm{int},\mathrm{pll}}i_dL_g/\mathrm{\omega}_{\mathrm{s}} }{1-k_{\mathrm{p},\mathrm{pll}}L_g/\mathrm{\omega}_{\mathrm{s}}\cdot i_d}\\
\end{aligned}
\label{ODE1A}
\end{equation}
\vspace{-0.5mm}
\begin{equation}
\begin{aligned}
f_{\mathrm{dvc},1}\left( \delta ,i_{\mathrm{dq}},\omega_{\mathrm{vdc}}{}_{\Delta} \nu _{\mathrm{dc}} \right) = & P_{in}-i_dU_g\cos \delta +i_qU_g\sin \delta  \\ & +\frac{C_{\mathrm{dc}}/\omega _s}{2}\mathrm{\zeta}_{\mathrm{dvc}}\omega _{\mathrm{dvc}}\varDelta \left( v_{\mathrm{dc}} \right) ^2
\\
f_{\mathrm{dvc},2}\left( \delta ,i_{\mathrm{dq}} \right) = & \frac{2}{C_{\mathrm{dc}}/\omega _s}\left( P_{in}-i_dU_g\cos \delta \right.\\ & \left. +i_qU_g\sin \delta \right)  \\
\end{aligned}
\label{ODE2A}
\end{equation}
\vspace{-0.5mm}
\begin{equation}
f_{\mathrm{tvc}}\left( \delta, i_q \right) = -i_q+k_vU_g\cos \delta -k_vi_qX_g-k_vV_{ref}
\label{ODE3A}
\end{equation}
\label{ODEA}
\end{subequations}
\normalsize

\section{Reduced-Order Model Derivation} \label{B}
\subsection{Bandwidth sequence: \texorpdfstring{$\omega_{pll} \text{ larger than } \omega_{dvc}$}{}}
\begin{figure}[t!]
\centering
\includegraphics[width=2.1in]{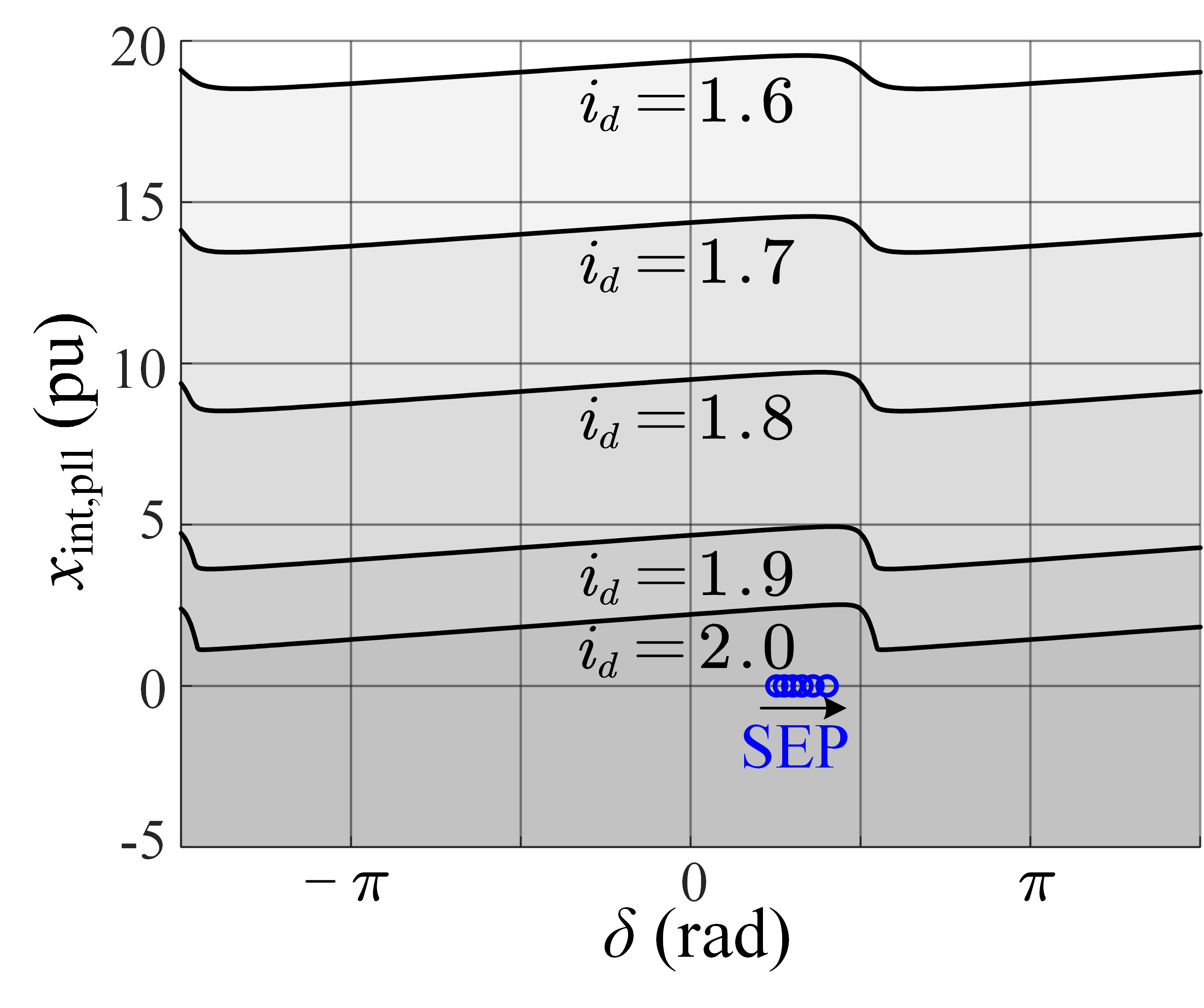}
\caption{Stability region of PLL under different $i_d$ ($\omega_{\mathrm{pll}}=15~\mathrm{Hz}\times2\pi$).}
\label{PLLSR}
\end{figure}

The PLL \eqref{PLL} is considered as fast subsystem, and defining $\varepsilon =\frac{1}{\omega _{\mathrm{pll}}}\cdot \left( 1-k_{\mathrm{p},\mathrm{pll}}L_g/\mathrm{\omega}_{\mathrm{s}}\cdot i_d \right)$, It clear that as $\omega _{\mathrm{pll}}\rightarrow \infty $, $\varepsilon\rightarrow0$. Introducing the fast time-scale $\tau=t/\varepsilon$, the fast subsystem equations become:
\small
\begin{equation}
    \left\{ \begin{array}{l}
	\frac{d}{d\tau}\delta =(X_gi_d-U_g\sin \delta )+\frac{x_{\mathrm{int}\_pll}}{\omega _{pll}}\\
	\frac{d}{d\tau}x_{\mathrm{int},\mathrm{pll}}=\zeta _{\mathrm{pll}}(X_gi_d-U_g\sin \delta +x_{\mathrm{int},\mathrm{pll}}L_g/\omega _s\cdot i_d)\\
\end{array} \right. 
\end{equation}
\normalsize
Since $\omega_{\mathrm{pll}}$ is large, $x_{\mathrm{int,pll}}/\omega_{\mathrm{pll}} \approx 0$. Under a fixed $i_d$, this subsystem rapidly converges to a SEP defined by:
\small
\begin{equation}
    \bar{\delta}=\mathrm{asin} \left( \frac{X_gi_d}{U_g} \right), \bar{x}_{\mathrm{int,pll}}=0
    \label{deltaf_sep}
\end{equation}
\normalsize
The SEP is uniformly exponentially stable if the following conditions hold:
\small
\begin{equation}
        i_d<\frac{\sqrt{{U_g}^2-\left( X_gi_d \right) ^2}}{\zeta _{\mathrm{pll}}L_g/\omega _s}, i_d<U_g/X_g, i_d<1/\omega _{\mathrm{pll}}L_g
\label{idreq}
\end{equation}
\normalsize
And the stability region for different $i_d$ is illustrated in \figref{PLLSR}
Following singular perturbation theory, we represent the fast dynamics as:
\small
\begin{equation}
\delta \left( t \right) =\bar{\delta}\left( t \right) +\delta ^*\left( \tau \right) +O\left( \varepsilon \right) 
\end{equation}
\normalsize
where the boundary-layer dynamics satisfy:
\small
\begin{equation}
\frac{d}{d\tau}\delta ^*\left( \tau \right) =X_gi_d-U_g\sin \left( \mathrm{asin} \left( \frac{X_gi_d}{U_g} \right) +\delta ^*\left( \tau \right) \right) 
\end{equation}
\normalsize
The DVC loop is the slow subsystem. Following the asymptotic expansion methodology of O’Malley and Smith \cite{smith1985singular, o1988nonlinear}, the slow variables are expanded as:  
\small
\begin{equation}
    i_d=\bar{i}_d+\varepsilon i_d^*\left( \tau \right) +O\left( \varepsilon \right) 
\end{equation}
\normalsize
where $\bar{i}_d$ and $\overline{\varDelta \left( v_{\mathrm{dc}} \right) ^2}$ satisfy the reduced-order slow subsystem equations:
\small
\begin{equation}
\frac{d}{dt}\left[ \begin{array}{c}
	\bar{i}_d\\
	\overline{\varDelta \left( v_{\mathrm{dc}} \right) ^2}\\
\end{array} \right] =\left[ \begin{array}{c}
	\omega _{\mathrm{dvc}}f_{\mathrm{dvc},1}\left( \bar{\delta},i_{\mathrm{dq}},\omega _{\mathrm{dvc}}\overline{\varDelta \left( v_{\mathrm{dc}} \right) ^2} \right)\\
	f_{\mathrm{dvc},2}\left( \bar{\delta},i_{\mathrm{dq}} \right)\\
\end{array} \right] 
\label{DVCs_D}
\end{equation}
\normalsize
with $\bar{\delta}=\mathrm{asin} \left( \frac{X_gi_d}{U_g} \right)$ substituted from the equilibrium of the fast subsystem.

The boundary layer correction terms $\varepsilon i_d^*\left( \tau \right)$ and $\varepsilon \left[ \varDelta \left( v_{\mathrm{dc}} \right) ^2 \right] ^*$ are small terms $O(\varepsilon)$ that rapidly decay and thus do not affect the reduced-order slow-subsystem differential equations. They serve only as initial value corrections, representing the transient effect of boundary-layer dynamics on the slow subsystem immediately after the fault. These corrections are approximated as:
\small
\begin{subequations}
\begin{equation}
\begin{aligned}
	{i_d}^*\left( \tau \right) &=-\omega _{\mathrm{dvc}}\int_{\tau}^{\infty}{f_{\mathrm{dvc},1}\left( \bar{\delta}(0)+\delta ^*(\tau ),i_{\mathrm{dq}}(0),\omega _{\mathrm{dvc}}\varDelta \left( v_{\mathrm{dc}} \right) ^2(0) \right)}\\
	&\quad -f_{\mathrm{dvc},1}\left( \bar{\delta}(0),i_{\mathrm{dq}}(0),\omega _{\mathrm{dvc}}\varDelta \left( v_{\mathrm{dc}} \right) ^2(0) \right) d\tau\\
	&\approx k_{\mathrm{p},\mathrm{dvc}}\frac{2}{C_{\mathrm{dc}}}U_gi_d(0)\sin \bar{\delta}(0)\cdot \frac{\delta (0)-\bar{\delta}(0)}{U_g\cos \bar{\delta}(0)}e^{-U_g\cos \bar{\delta}(0)\tau}\\
\end{aligned}
\end{equation}
\begin{equation}
\begin{aligned}
	\left[ \varDelta \left( v_{\mathrm{dc}} \right) ^2 \right] ^*\left( \tau \right) &=-\int_{\tau}^{\infty}{f_{\mathrm{DVC},2}\left( \bar{\delta}\left( 0 \right) +\delta ^*\left( \tau \right) ,i_{\mathrm{dq}}\left( 0 \right) ,\omega _{\mathrm{dvc}}\varDelta \left( v_{\mathrm{dc}} \right) ^2\left( 0 \right) \right)}\\
	&\quad -f_{\mathrm{DVC},2}\left( \bar{\delta}\left( 0 \right) ,i_{\mathrm{dq}}\left( 0 \right) ,\omega _{\mathrm{dvc}}\varDelta \left( v_{\mathrm{dc}} \right) ^2\left( 0 \right) \right) d\tau\\
	&\approx \frac{2}{C_{\mathrm{dc}}}U_gi_d\left( 0 \right) \sin \bar{\delta}\left( 0 \right) \cdot \frac{\delta \left( 0 \right) -\bar{\delta}\left( 0 \right)}{U_g\cos \bar{\delta}\left( 0 \right)}e^{-U_g\cos \bar{\delta}\left( 0 \right) \cdot \tau}\\
\end{aligned}
\end{equation}
\label{initialvaluedvc}
\end{subequations}
\normalsize
where $\delta(0)$ denotes the initial post-fault value, differing from $\bar{\delta}(0)$ which determined by the initial value of slow variables.

\subsection{Bandwidth sequence: \texorpdfstring{$\omega_{dvc} \text{ larger than } \omega_{pll}$}{}}
The DVC \eqref{DVC} is considered as fast subsystem, and defining $\varepsilon = 1/\omega_{dvc}$, It clear that as $\omega _{\mathrm{dvc}}\rightarrow \infty $, $\varepsilon\rightarrow0$. Introducing the fast time-scale $\tau=t/\varepsilon$, and do variable substitution $\sigma =\omega _{\mathrm{dvc}}\cdot \varDelta \left( v_{\mathrm{dc}} \right) ^2$, the fast subsystem equations become:
\small
\begin{equation}
    \begin{cases}
	\frac{d}{d\tau}i_d=\left( P_{in}-i_dU_g\cos \delta +i_qU_g\sin \delta \right) +\frac{C_{\mathrm{dc}}/\omega _s}{2}\mathrm{\zeta}_{\mathrm{dvc}}\sigma\\
	\frac{d}{d\tau}\sigma =\frac{2}{C_{\mathrm{dc}}}\left( P_{in}-i_dU_g\cos \delta +i_qU_g\sin \delta \right)\\
\end{cases}
\end{equation}
\normalsize
It is a linear differential equation for fixed $\delta$, and it is easy to prove that if $\mathrm{cos}\delta>0$, the Jacobian is negative, which means that the fast subsystem is globally asymptotic stable. They rapidly converges to $\bar{i}_d$ and $\bar{\sigma}$:
\small
\begin{equation}
P_{in}-\bar{i}_dU_g\cos \delta +i_qU_g\sin \delta =0, \bar{\sigma}=0
\end{equation}
\normalsize
Similarly, following singular perturbation theory, the fast-subsystem dynamics can be expressed as:
\small
\begin{equation}
\left\{ \begin{array}{l}
	i_d\left( t \right) =\bar{i}_d\left( t \right) +i_d\left( \tau \right) +O\left( \varepsilon \right)\\
	\sigma \left( t \right) =\bar{\sigma}\left( t \right) +\sigma \left( \tau \right) +O\left( \varepsilon \right)\\
\end{array} \right. 
\end{equation}
\normalsize
where the boundary-layer dynamics satisfy:
\small
\begin{equation}
\left\{ \begin{array}{l}
	\frac{d}{d\tau}{i_d}^*\left( \tau \right) =-{i_d}^*\left( \tau \right) U_g\cos \delta \left( 0 \right) +\frac{C_{\mathrm{dc}}/\omega _s}{2}\mathrm{\zeta}_{\mathrm{dvc}}\sigma ^*\left( \tau \right)\\
	\frac{d}{d\tau}\sigma ^*\left( \tau \right) =-\frac{2}{C_{\mathrm{dc}}}{i_d}^*\left( \tau \right) U_g\cos \delta \left( 0 \right)\\
\end{array} \right.
\end{equation}
\normalsize
The PLL loop is the slow subsystem. Following the asymptotic expansion methodology \cite{o1988nonlinear, smith1985singular} the slow variables are expanded as:
\small
\begin{equation}
\begin{cases}
	\delta \left( t \right) =\overline{\delta }\left( t \right) +\varepsilon \delta ^*\left( \tau \right) +O\left( \varepsilon \right)\\
	x_{\mathrm{int},\mathrm{pll}}\left( t \right) =\bar{x}_{\mathrm{int},\mathrm{pll}}\left( t \right) +\varepsilon x_{\mathrm{int},\mathrm{pll}}^{*}\left( \tau \right) +O\left( \varepsilon \right)\\
\end{cases}
\end{equation}
\normalsize
where $\overline{\delta }\left( t \right)$ and $\bar{x}_{\mathrm{int},\mathrm{pll}}$ satisfy the reduced-order slow subsystem equations:
\begin{equation}
\left[ \begin{array}{c}
	\dot{\delta}\\
	\dot{x}_{\mathrm{int},\mathrm{pll}}\\
\end{array} \right] =\left[ \begin{array}{c}
	\omega _{\mathrm{pll}}f_{\mathrm{pll},1}\left( \delta ,\bar{i}_d,\frac{x_{\mathrm{int},\mathrm{pll}}}{\omega _{\mathrm{pll}}} \right)\\
	\omega _{\mathrm{pll}}f_{\mathrm{pll},2}\left( \delta ,\bar{i}_d,x_{\mathrm{int},\mathrm{pll}} \right)\\
\end{array} \right] 
\end{equation}
with $\bar{i}_d$ substituted from the equilibrium of the fast subsystem.

The boundary layer correction terms $\varepsilon\delta^*\left( \tau \right)$ and $\varepsilon x_{\mathrm{int,pll}}^*\left( \tau \right)$ are small terms $O(\varepsilon)$ that rapidly decay and thus do not affect the reduced-order slow-subsystem differential equations. They serve only as initial value corrections, representing the transient effect of boundary-layer dynamics on the slow subsystem immediately after the fault. These corrections are approximated as:
\small
\begin{equation}
\begin{aligned}
    \delta ^*\left( \tau \right) =&-\omega _{\mathrm{pll}}\int_{\tau}^{\infty}{f_{\mathrm{pll},1}\left( \delta \left( 0 \right) ,i_{\mathrm{d}}^{*},\frac{x_{\mathrm{int},\mathrm{pll}}\left( 0 \right)}{\omega _{\mathrm{pll}}} \right) d\tau}
\\
x_{\mathrm{int},\mathrm{pll}}^{*}\left( \tau \right) =&-\omega _{\mathrm{pll}}\int_{\tau}^{\infty}{f_{\mathrm{pll},2}\left( \delta \left( 0 \right) ,i_{\mathrm{d}}^{*},x_{\mathrm{int},\mathrm{pll}}\left( 0 \right) \right) d\tau}
\end{aligned}
\end{equation}
\normalsize

\subsection{Stability analysis of TVC subsystem}
Since the DVC loop does not affect the TVC loop, only the influence of the PLL is considered. The differential equation \eqref{TVC}, which governs $i_q$, is a linear system with $\delta$ serving as the input. It is globally exponentially stable uniformly in $\delta$, and therefore input-to-state stable (ISS) according to Lemma 4.6 in \cite{nonlinearbook_Khalil}. In particular, if $\delta(t)$ converges to a constant, then $i_q(t)$ asymptotically converges to the corresponding steady-state value. In summary, if $\delta$ can be stable, then \eqref{TVC} is stable.

\section{System Parameters}
If not specified, the system parameters used in this paper are listed in \tabref{Parameter}. The simulations and all numerical calculations in this paper are also available online at \cite{Online}.

\begin{table}[h]
\centering
\begin{threeparttable}
\caption{Parameters in Simulation and HIL Experiment}\label{Parameter}
\renewcommand{\arraystretch}{1.2} 
\begin{tabular}{@{\hspace{0.2em}} l @{\hspace{0.4 em}} l @{\hspace{0.3 em}} l @{\hspace{0.2 em}}}
\hline \hline
\multirow{2}{*}{\raggedright Parameters} & \multicolumn{2}{@{\hspace{-0.1em}}l}{Value\tnote{1}} \\
 \cline{2-3}
& Simulation & HIL Experiment \\
\hline
Base capacity $S_b$                         &  $1$ &  $1\,\text{MVA}$  \\
Base voltage $V_b$                          &  $1$ &  $690\,\text{V}$  \\
Base current $I_b$                          &  $1$ &  $1.45\,\text{kA}$  \\
Fundamental frequency $\omega_s/2\pi$              &  $50\,\text{Hz}$ &  $50\,\text{Hz}$  \\
Grid voltage $U_g$                          &  $1$~pu &  $690\,\text{V}$  \\
Input power $P_{\text{in}}$                        &  $1$~pu &  $1\,\text{MW}$ \\
Short circuit ratio (SCR)                   &  $2.1$ &  $2.1$  \\
Grid impedance $Z_g$                        & \scriptsize $0.47j+0.002$~pu & \makecell[l]{ \scriptsize $0.72\text{mH}+0.95\text{m}\Omega$} \\
DC voltage reference $V_{\text{dc,ref}}$           &  $2.5$~pu &  $1725\,\text{V}$ \\
DC capacitor $C_{\text{dc}}$                &  $12.5$~pu &  $83\,\text{mF}$ \\
DC voltage limiter $V_{\text{dc,limit}}$    &  $2$-$3$~pu ($\pm20\%$) &  $1380$ - $2070\,\text{V}$ \\
PLL PI controller                           & \makecell[l]{ \scriptsize $k_{\text{p,pll}} = \omega_{\text{pll}}$ \\ \scriptsize $k_{\text{i,pll}} = \zeta_{\text{pll}} \cdot  k_{\text{p,pll}}$, $\zeta_{\text{pll}}=1/4$} &same   \\
DVC PI controller                           & \makecell[l]{  \scriptsize $k_{\text{p,dvc}}=\omega_{\text{dvc}}\cdot C_{\text{dc}}/2\omega_s$  \\ \scriptsize $k_{\text{i,dvc}} = k_{\text{p,dvc}}\cdot\omega_{\text{dvc}}/4$} & same \\
TVC droop coefficient $k_v$                       & $2$~pu & same   \\
TVC voltage reference $V_{\text{t,ref}}$           & $0.81$~pu & $560\,\text{V}$   \\
\makecell[l]{Inner current control \\ loop bandwidth}        & $1\,\text{kHz}$ &same   \\
Current limit\tnote{2} \ $I_{\text{limit}}$        & $2.5$~pu  & 3.1~kA   \\
LCL filter $L_f$                            & $0.05$~pu & $75.7\,\mu\text{H}$  \\
LCL filter $C_f$                            & $0.01$~pu & $66.8\,\mu\text{F}$  \\
Switching frequency                         & averaging model & $10$~kHz  \\
DSP control frequency                       & $-$ & $10$~kHz  \\
\hline \hline
\end{tabular}
\begin{tablenotes}
\footnotesize
\scriptsize
\item[1] The pu values in simulation and experiment are the same.
\item[2] Rectangular current limiter.
\end{tablenotes}
\end{threeparttable}
\end{table}


\ifCLASSOPTIONcaptionsoff
  \newpage
\fi
\bibliographystyle{IEEEtran}
\bibliography{References}

\end{document}